\documentclass[12pt,a4paper]{article}

\RequirePackage[OT1]{fontenc}
\RequirePackage{amsthm}
\RequirePackage[cmex10]{amsmath}
\RequirePackage{natbib}
\RequirePackage[colorlinks,citecolor=blue,urlcolor=blue, linkcolor=blue, bookmarksopen=true, breaklinks]{hyperref}

\usepackage[margin=1in]{geometry}
\usepackage[latin1]{inputenc}
\usepackage{amsmath}
\usepackage{amsfonts}
\usepackage{amssymb}
\usepackage{comment}
\usepackage{bbm}
\usepackage{graphicx}
\usepackage[usenames,dvipsnames]{color}
\usepackage{multirow}
\usepackage[labelfont=bf]{caption}
\usepackage{placeins}
\usepackage{array}
\usepackage{enumitem}
\usepackage{rotating}
\usepackage{adjustbox}
\usepackage{mathrsfs}
\usepackage{gensymb}
\usepackage{float}
\usepackage{subcaption}
\usepackage{wasysym}
\usepackage[linewidth=1pt]{mdframed}
\usepackage{stmaryrd}
\usepackage{thmtools}
\usepackage{thm-restate}
\usepackage{cleveref}
\usepackage{tikz,subcaption}

\theoremstyle{definition}

\theoremstyle{plain}

\newtheoremstyle{mytheoremstyle}
    {\topsep}                    
    {\topsep}                    
    {\itshape}                   
    {}                           
    {\scshape}                   
    {:}                         
    {.5em}                       
    {} 
\theoremstyle{mytheoremstyle}

\title{\textsc{Patience ensures fairness}\protect}

\author{\textsc{Florian Brandl\thanks{%
Department of Economics, University of Bonn, Bonn, Germany. Email: florian.brandl@uni-bonn.de} and Andrew Mackenzie\thanks{%
Department of Economics, Rutgers University, New Brunswick, NJ, United States of America. Email: andrew.k.mackenzie@rutgers.edu} \hspace{0.01mm}} \thanks{Florian Brandl acknowledges support by the DFG under the Excellence Strategy EXC-2047. We thank Vilmos Komornik and seminar participants of the Randomization, Neutrality, and Fairness Workshop at the Simons Laufer Mathematical Sciences Institute (Berkeley, October 2023) and the COMSOC Seminar Series (Online, April 2024).}}
\date{This draft: \today}

\begin{document}

\maketitle

\begin{abstract}
We revisit the problem of fairly allocating a sequence of time slots when agents may have different levels of patience \citep{Mackenzie-Komornik2023}. For each number of agents, we provide a lower threshold and an upper threshold on the level of patience such that (i)~if each agent is at least as patient as the lower threshold, then there is a {\it proportional} allocation, and (ii)~if each agent is at least as patient as the upper threshold and moreover has weak preference for earlier time slots, then there is an {\it envy-free} allocation. In both cases, the proof is constructive. \\


\end{abstract}

\hypertarget{Section1}{}
\section{Introduction}

\hypertarget{Section1.1}{}
\subsection{Overview}

Can fairness be achieved without dividing goods into pieces, redistributing money, or randomizing, simply by taking turns? For example, can siblings identify a fair schedule for sharing an inherited estate, can timeshare owners fairly allocate calendar dates, or can divorced parents identify a fair custody schedule? If all parties are sufficiently impatient, then clearly the answer is no. In this article, we revisit a recent stylized model of taking turns \citep{Mackenzie-Komornik2023}, and we prove that if all parties are sufficiently patient, then the answer is yes. Moreover, if all parties are sufficiently patient and have weak preference for earlier time slots, then the answer is emphatically yes.

Crucially, we consider notions of fairness that make no reference to property rights. Indeed, a natural interpretation of our problem is that some commonly owned property must be privatized, and we seek a fair allocation to {\it define} endowments. For example, (i)~in the inheritance scenario, the will specifies only that the estate passes to the decedent's children, (ii)~in the timeshare scenario, the owners contribute equally to bidding on a property without first signing a contract detailing how dates will be allocated should they win, and (iii)~in the custody scenario, the child belongs to a partnership that dissolves.

In the model, the {\it time slots} in $T = \{1, 2, \dots\}$ are to be partitioned into {\it schedules} for~$n$ agents, and each agent $i$ has preferences over schedules that can be represented by a countably additive probability measure $u_i$. An allocation is {\it proportional} if each agent measures his own schedule to be worth at least $\frac{1}{n}$ \citep{Steinhaus1948}, and is moreover {\it envy-free} if no agent measures another's schedule to be worth more than his own (\citealp{Tinbergen1946}; \citealp{Foley1967}). These fairness notions are central in the large literature on fair division, which spans perfectly divisible cakes, collections of indivisible objects, and classical exchange economies.

We formalize each agent's level of patience using a generalization of the heavy-tails condition of Kakeya (\citealp{Kakeya1914}; \citealp{Kakeya1914b}): for each $k \in [0, \infty)$, we say that $u_i$ is {\it $k$-Kakeya} if it always measures the relative value of the future to the present to be at least $k$, or equivalently, if for each $t \in T$ we have $u_i(\{t+1, t+2, \dots\}) \geq k \cdot u_i(\{t\})$. We let $\mathcal{U}_k$ denote the set of these utility functions, and remark that this is a general class that includes many standard models of discounting; for example, if $u_i$ is geometric discounting with respect to discount factor $\delta_i \in (0, 1)$, in the sense that for each $S \subseteq T$ we have $u_i(S) = (1-\delta_i) \sum_{t \in S} \delta_i^{t-1}$, then $u_i \in \mathcal{U}_k$ if and only if $\delta_i \geq \frac{k}{k+1}$. In addition, we sometimes impose weak preference for earlier time slots: we say that $u_i$ is {\it monotonic} if it always assigns an earlier time slot at least as much value as a later time slot, and we let $\mathcal{U}_{\text{M}}$ denote the set of these utility functions.

Our main results guarantee that there are fair allocations so long as the agents are sufficiently patient, given the number of agents~$n$. In particular, 
\begin{itemize}
\item if all agents have utility functions in $\mathcal{U}_{2n-3}$, then there are proportional allocations (\hyperlink{Theorem1}{Theorem~1}), and

\item if all agents have utility functions in $\mathcal{U}_{2 \cdot 3^{[2 \cdot 32^n \cdot n^7 \cdot (n!)^2-1]}-1} \cap \mathcal{U}_{\text{M}}$, then there are envy-free allocations (\hyperlink{Corollary4}{Corollary~4}, a corollary of \hyperlink{Theorem2}{Theorem~2}).
\end{itemize}
Under the interpretation that the interval between time slots is fixed, such as one day or one month, \hyperlink{Corollary4}{Corollary~4} requires a finite but astronomical amount of patience to guarantee the existence of an envy-free allocation. If instead the interpretation is that we are partitioning continuous time $[0, \infty)$ into indivisible time slots of the form $[t, t+\Delta)$ and then allocating the resulting time slots, for example when each agent has an exponential density and thus a geometric utility function over the resulting time slots, then the requirement is that the discretization should be sufficiently fine: there are envy-free allocations for any sufficiently small $\Delta>0$.

Notably, both proofs are constructive. The former result is established using \textsc{Iterative Cycle Apportionment}, a novel modification of the \textsc{Iterative Apportionment} procedure due to \cite{Mackenzie-Komornik2023}. For both procedures, iteratively the remaining agents construct and assign a schedule, but only in the novel procedure do they automatically skip certain time slots in earlier stages to reserve them for later stages. The latter result is established using the \textsc{Tripartition Algorithm}, a novel alternative to the classic \textsc{Greedy Algorithm} \citep{Renyi1957}. For both procedures, an agent $i$ is given a schedule $S$ that he considers `very divisible' and a target value $v \in (0, u_i(S))$, and uses the procedure to output a schedule $S^* \subseteq S$ such that (i)~$u_i(S^*) = v$, and (ii)~$i$ considers $S \setminus S^*$ `somewhat divisible,' but only with the novel procedure does consensus across all agents that $S$ is `very divisible' lead to consensus across all agents that {\it both} $S^*$ and $S \setminus S^*$ are `somewhat divisible.' Ultimately, the \textsc{Tripartition Algorithm} allows us to import certain bounded envy-free protocols for dividing a perfectly divisible cake into our setting, and the bound given above follows from applying this technique to the \textsc{Psomas} protocol \citep{Psomas2026}; we also demonstrate this technique using \textsc{Divide and Choose} for $n=2$ and the \textsc{Selfridge-Conway} protocol (discovered independently by John Selfridge and John Conway; see \citealp{Brams-Taylor1995}) for $n=3$.

Altogether, our main results formalize the notion that if all parties are sufficiently patient, then fairness can be achieved without dividing goods into pieces, redistributing money, or randomizing, simply by taking turns. In particular, this statement holds with no further hypothesis for the weaker fairness notion of proportionality, and holds under the additional hypothesis that all parties have weak preference for earlier time slots for the stronger fairness notion of no-envy.

\hypertarget{Section1.2}{}
\subsection{Related literature}

We organize this discussion into four topics: (i)~previous results about fairly taking turns, (ii)~the continuous time benchmark, (iii)~related results for the case of identical preferences, and (iv)~related results about Fisher markets.

\vspace{\baselineskip} \noindent \textsc{Previous results.} For the problem we consider---fairly taking turns when agents need not have identical preferences---\cite{Mackenzie-Komornik2023} established that
\begin{itemize}
\item if $n=2$ and both agents have utility functions in $\mathcal{U}_{n-1}$, then envy-free allocations can be constructed using a version of the ancient \textsc{Divide and Choose} procedure;

\item if $n=3$ and all agents have utility functions in $\mathcal{U}_{n-1}$, then proportional allocations can be constructed using either their \textsc{Iterative Apportionment} procedure or their \textsc{Simultaneous Apportionment} procedure; and

\item in general, for each $\varepsilon > 0$, if all agents have utility functions in $\mathcal{U}_{\frac{1-\varepsilon}{\varepsilon}} \cap \mathcal{U}_{\text{M}}$, then allocations that are $\varepsilon$-approximately envy-free can be constructed using a simple \textsc{Round-Robin} procedure \citep{Caragiannis-Kurokawa-Moulin-Procaccia-Shah-Wang2019}.
\end{itemize}
Observe that there may not be proportional allocations if the agents do not have utility functions in $\mathcal{U}_{n-1}$, as in this case the first time slot alone may be worth more than~$\frac{1}{n}$ to everyone; thus (i)~when the number of agents is small, these results provide a tight bound on the level of patience required to guarantee that there are exactly fair allocations, and (ii)~in general, these results require a level of patience above what may strictly be necessary in order to guarantee that there are approximately fair allocations. Our contribution is to require a level of patience above what may strictly be necessary in order to guarantee that there are exactly fair allocations in general.

\vspace{\baselineskip} \noindent \textsc{Continuous time.} There is a recurring theme in economics that parameterized discrete models often approximate continuous models, and this is the case here. Indeed, increasing the patience of the agents is akin to decreasing the length of time between time slots, for which the ``limit" is the continuous model with $T = [0, \infty)$ and preferences represented by atomless probability measures that assign positive weight to all tails, and intuitively these utility functions capture infinite patience because for each $t \in T$, the relative value of the present to the future, $\frac{u_i(\{t\})}{u_i((t, \infty))}$, is zero. This in turn is a special case of the classic problem of dividing a perfectly divisible cake \citep{Steinhaus1948}, for which it is well-known that there are envy-free allocations \citep{Dubins-Spanier1961}. Provided that agents have weak preference for earlier time slots, it's easy to show that there are approximately fair allocations for which the margin of error vanishes as the agents grow arbitrarily patient: simply put the agents in a line and iteratively (i)~assign the current time slot to the agent at the front of the line, and (ii)~send that agent to the back of the line.\footnote{In addition to constructing allocations that are approximately fair, this \textsc{Round-Robin} procedure \citep{Caragiannis-Kurokawa-Moulin-Procaccia-Shah-Wang2019} has many practical advantages. For example, it incentivizes agents to honestly report their preferences when this is private information, and it can assign each time slot without asking agents to report their values for future time slots. Unfortunately, however, even for two agents with a restricted preference domain, each of these practical advantages is incompatible with (exact) proportionality \citep{Mackenzie-Komornik2023}.} Our results go further: deviating from previous work on fair allocation in purely discrete models, we pursue the existence of {\it exactly} fair allocations. In this way, we provide a simple example of a parameterized discrete model that, under economically meaningful conditions on its parameters, behaves {\it exactly} like its continuous limit.

\vspace{\baselineskip} \noindent \textsc{Identical preferences.} In the special case that $n$ agents have identical preferences given by a common discount factor $\delta \in [\frac{n-1}{n}, 1)$, the existence of envy-free allocations follows from a classic result for repeated games: if the agents are sufficiently patient given the number of action profiles, then for each convex combination of stage game payoffs, there is a sequence of action profiles that generates those payoffs (\citealp{Sorin1986}; \citealp{Fudenberg-Maskin1991}). Indeed, consider the repeated game with the following stage game: the oldest agent selects any agent, every other agent has a single dummy action, and the selected agent receives payoff~$1$ while the others receive payoff $0$. By the classic result, there is a sequence of selected agents for which each agent receives payoff~$\frac{1}{n}$, and as agents have identical preferences, the associated allocation is envy-free.

There are a variety of procedures that yield constructive proofs of the classic result. Translated into fair division, (i)~the \textsc{Fudenberg-Maskin} procedure \citep{Fudenberg-Maskin1991} iteratively selects any agent whose cumulative utility is lowest, (ii)~the more flexible \textsc{Sorin} procedure \citep{Sorin1986} iteratively selects any agent who still requires at least $\frac{1}{n}$ of the remaining utility, and (iii)~the iterative application of R\'{e}nyi's \textsc{Greedy Algorithm} \citep{Renyi1957} iteratively asks each agent to construct a schedule for himself that achieves his target utility, by iteratively taking each remaining time slot unless it would make his schedule too valuable. All of these procedures, and many others, work more generally whenever agents have identical $(n-1)$-Kakeya preferences, and it follows that in our model, proportionality is equivalent to guaranteeing each agent his {\it maximin share} (\citealp{Hill1987}; \citealp{Budish2011}): the maximum utility that could simultaneously be guaranteed to everybody in the economy where all the other agents are replaced by copies of the given agent. That said, none of these procedures works when preferences are not identical \citep{Mackenzie-Komornik2023}. This reinforces a recurring theme in the literature: allowing agents to have different levels of patience---and thereby introducing the possibility of gains from intertemporal trade---often yields technical challenges and economic implications.\footnote{For example, allowing one agent to be more patient than another has significant implications for capital accumulation (\citealp{Ramsey1928}; \citealp{Becker1980}; \citealp{Rader1981}), bargaining \citep{Rubinstein1982}, reputation \citep{Fudenberg-Levine1989}, repeated games (\citealp{Lehrer-Pauzner1999}; \citealp{Salonen-Vartiainen2008}; \citealp{Chen-Takahashi2012}; \citealp{Sugaya2015}), preference aggregation toward a social discount factor (\citealp{Weitzman2001}; \citealp{Jackson-Yariv2015}; \citealp{Chambers-Echenique2018}), and endogenous discounting \citep{Kochov-Song2023}.}

\vspace{\baselineskip} \noindent \textsc{Fisher markets.} It is tempting to seek proofs of our theorems that use only existing techniques from integer programming, and though we are not aware of such a proof, there are some intriguing related results in this direction concerning Fisher markets. Intuitively, the idea is to allow randomization in the model and impose risk neutrality, but then forbid randomization with integer constraints. From here, if we relax the integer constraints and assign monetary budgets (that cannot be consumed) to the agents, then we have a Fisher market with an infinite collection of goods, and there are competitive equilibria that can be calculated using the convex program of \cite{Eisenberg-Gale1959}. To pursue fairness, we equate the budgets, in which case the convex program maximizes the Nash product to compute a competitive equilibrium from equal incomes, and this is not only envy-free but moreover efficient. Unfortunately, this allocation will generally violate the integer constraints, or involve randomization, and it is trivial to obtain exact fairness ex-ante with randomization: simply select one of the agents uniformly at random to receive all of the time slots.

Remarkably, it is always possible to perturb the monetary budgets ``a small amount" and thereby reach a second Fisher market for which some competitive equilibrium satisfies the integer constraints \citep{Barman-Krishnamurthy2019}. Unfortunately, this does not help us find exactly fair allocations: when one agent's budget is slightly larger than another's, the resulting competitive equilibrium need not be envy-free. Even so, this is an intriguing approach to approximate fairness in our model.

\hypertarget{Section2}{}
\section{Model}

This section is organized into two subsections. In \hyperlink{Section2.1}{Section~2.1}, we introduce the formal definitions that are used to state our main results, while in \hyperlink{Section2.2}{Section~2.2}, we introduce some additional concepts that prove useful for our analysis.

Before proceeding, we make three brief remarks about notation. First, we distinguish the set of natural numbers $\mathbb{N}$ from the set of all time slots $T$ whenever the former serves a distinct purpose in the analysis from the latter, even though these two sets are equivalent. Second, we use $i$ to index agents, $t$ to index time slots, $r$ to index earliness {\it ranks}, and $\ell$ to index period {\it lengths}; ranks and lengths are introduced in \hyperlink{Section2.2}{Section~2.2}. Finally, we use integer interval notation: for each pair $a, b \in \mathbb{Z}$, we let $\llbracket a, b \rrbracket$ denote $\{a, a+1, \dots, b\}$.

\hypertarget{Section2.1}{}
\subsection{Concepts for stating our results}

In our model, we consider economies where a countably infinite collection of time slots is to be partitioned into schedules for the agents, and each agent has preferences over schedules that may be represented by a countably additive probability measure.

\vspace{\baselineskip} \noindent \textsc{Definition 1:} An {\it economy} is specified by a pair $(n, u)$ with $n \in \mathbb{N}$ as follows:
\begin{itemize}
\item $N \equiv \{1, 2, \dots, n\}$ is the set of {\it agents};

\item $T \equiv \{1, 2, \dots\}$ is the countably infinite collection of {\it time slots};

\item $\mathcal{S} \equiv 2^T$ is the collection of {\it schedules}, which is each agent's consumption space;

\item $u = (u_i)_{i \in N}$ is the profile of {\it utility functions}: for each $i \in N$, $u_i:\mathcal{S} \to [0,1]$ is a countably additive probability measure\footnote{That is, we have (i)~$u_i(T) = 1$, and (ii)~for each $S \in \mathcal{S}$, $u_i(S) = \sum_{t \in S} u_i(\{t\})$.} representing the preferences of $i$; and

\item $\Pi \subseteq \mathcal{S}^N$ is the collection of {\it (partitional) allocations}: for each $\pi \in \mathcal{S}^N$, we have $\pi \in \Pi$ if and only if (i)~for each pair of distinct $i, j \in N$, $\pi_i \cap \pi_j = \emptyset$, and (ii)~$\cup_{i \in N} \pi_i = T$.
\end{itemize}
Whenever we refer to an arbitrary economy, we assume all of the above notation.

\vspace{\baselineskip} We are interested in the existence of allocations that satisfy the following normative axioms, and in constructing these allocations whenever possible.

\vspace{\baselineskip} \noindent \textsc{Definition 2:} Fix an economy and let $\pi \in \Pi$. We say that $\pi$ satisfies
\begin{itemize}
\item {\it proportionality} if for each $i \in N$, $u_i(\pi_i) \geq \frac{1}{n}$;

\item {\it no-envy} if for each pair $i, j \in N$, $u_i(\pi_i) \geq u_i(\pi_j)$; and

\item {\it (Pareto) efficiency} if there is no $\pi' \in \Pi$ such that

\hspace{5mm} (i)~for each $i \in N$, $u_i(\pi'_i) \geq u_i(\pi_i)$, and

\hspace{5mm} (ii)~for some $i \in N$, $u_i(\pi'_i) > u_i(\pi_i)$.
\end{itemize}
We say that $\pi$ is {\it proportional}, {\it envy-free}, or {\it efficient}, accordingly.

\vspace{\baselineskip} It is easy to verify that no-envy implies proportionality. Our results involve economies where each agent is sufficiently patient given the number of agents, which we articulate using a generalization of a condition due to Kakeya (\citealp{Kakeya1914}; \citealp{Kakeya1914b}):

\vspace{\baselineskip} \noindent \textsc{Definition 3:} Fix an economy and a utility function $u_0$. For each $k \in [0, \infty)$, we say that $u_0$ is
\begin{itemize}
\item {\it $k$-Kakeya} if for each $t \in T$, we have $u_0(\{t+1, t+2, \dots\}) \geq k \cdot u_0 (\{t\})$; and

\item {\it monotonic} if $u_0(\{1\}) \geq u_0(\{2\}) \geq \dots$.
\end{itemize}
We let $\mathcal{U}_k \subseteq [0,1]^\mathcal{S}$ and $\mathcal{U}_{\text{M}}\subseteq [0,1]^\mathcal{S}$ denote the sets of $k$-Kakeya and monotonic utility functions, respectively.

\vspace{\baselineskip} Observe that an agent with a higher-level Kakeya utility function is, in a particular sense, more patient. At this point, we have introduced almost all of the formal definitions required to state our main results, with an exception: the statement of \hyperlink{Theorem2}{Theorem~2} involves {\it envy-free cellular cake protocols} and their {\it worst-case costs} for cake division problems. At a high level, these are query protocols that allow a mediator who initially does not know the agents' cake measures to ultimately construct an envy-free cake allocation, with costs measured by cut queries, and the formal definitions involve many details that are not related to the main ideas of this paper. See \hyperlink{Appendix4}{Appendix~4} for these formal definitions.

\hypertarget{Section2.2}{}
\subsection{Additional concepts for proving our results}

To begin, we introduce some simple methods for constructing a sub-schedule from an infinite source schedule.

\vspace{\baselineskip} \noindent \textsc{Definition 4:} Let $S \in \mathcal{S}$ be an infinite schedule, let $r \in \mathbb{N}$ be a rank, and let $\ell \in \mathbb{N}$ be a length. Then
\begin{itemize}
\item $\tau_r(S) \in S$ denotes the $r$th-earliest time slot in $S$, with singleton $S_r \equiv \{\tau_r(S)\}$;

\item $S_{[r)} \subseteq S$ denotes the {\it tail of $S$ starting at $r$}, $\cup_{r' \geq r} S_{r'} = \{ \tau_r(S), \tau_{r+1}(S), \dots \}$; and

\item $S_{[r|\ell)} \subseteq S$ denotes the {\it $\ell$-cycle of $S$ starting at~$r$} constructed by beginning with an empty basket, considering the time slots in $S_{[r)}$ in sequence, and iteratively adding one time slot and then skipping $\ell-1$ time slots.\footnote{To be precise using the language of modular arithmetic, if $R = \{1 + x \ell \mid x \in \{0, 1, \dots\} \}$ is the residue class of $1$ modulo $\ell$, then $S_{[r|\ell)} = \{\tau_{r'}(S_{[r)})\mid r'\in R\}$.}
\end{itemize}
For example, if $S = \{2, 4, \dots\}$ is the schedule of even time slots, then $S_{[3)} = \{6, 8, \dots\}$ is the tail of $S$ starting at the third even time slot, and $S_{[3|2)} = \{6, 10, 14, 18, \dots\}$ is formed from $S_{[3)}$ by iteratively adding one time slot in $S_{[3)}$ and skipping one time slot in $S_{[3)}$.

\vspace{\baselineskip} We discuss tails and cycles in sequence. First, a tail of a reference schedule~$S$ consists of an initial time slot and all future ones in $S$. Our notion of patience involves tails of $T$: a utility function is $k$-Kakeya if for each $r \in \mathbb{N}$, the value of the tail of~$T$ starting at~$r+1$ is at least $k$ times the value of $\{r\} = T_r$ alone. We will more generally be interested in other reference schedules with this property.

\vspace{\baselineskip} \noindent \textsc{Definition 5:} Fix an economy. For each $i \in N$, each $k \in [0, \infty)$, and each $S \in \mathcal{S}$, we say that $S$ is {\it $k$-divisible for $i$} if for each $t \in S$, we have $u_i(\{t' \in S \mid t' > t\}) \geq k \cdot u_i(\{t\})$. Observe that for each infinite schedule $S$, this is equivalent to the requirement that for each $r \in \mathbb{N}$, we have $u_i(S_{[r+1)}) \geq k \cdot u_i(S_r)$. Observe also that $u_i$ is $k$-Kakeya if and only if the set of all time slots $T$ is $k$-divisible for $i$.

\vspace{\baselineskip} This terminology is justified by the fact that if a schedule $S$ is $1$-divisible for $i$, then for each target value $v \in [0, u_i(S)]$, an adaptation of the well-known \textsc{Greedy Algorithm} for representing numbers in arbitrary bases \citep{Renyi1957} can be used to cut and remove a sub-schedule from $S$ that is worth precisely $v$. More generally, if $S$ is $k$-divisible for $k \in [1, \infty)$, then the \textsc{Greedy Algorithm} can successfully be used in this manner $\lfloor k \rfloor$ consecutive times, where $\lfloor k \rfloor$ denotes the greatest integer less than or equal to $k$.

\vspace{\baselineskip} \noindent \textsc{Definition 6:} {\it Greedy Algorithm.} Fix an economy. For each $i \in N$, each $S \in \mathcal{S}$, and each $v \in [0, u_i(S)]$, let $\mathcal{G}_i(v|S) \subseteq S$ denote the {\it greedy schedule (for agent $i$ given source $S$ and target $v$)} constructed by beginning with an empty basket, considering the time slots in~$S$ in sequence, and adding each time slot to the basket if and only if the value of the basket according to $u_i$ will not exceed $v$.

\hypertarget{TheoremMK}{}
\vspace{\baselineskip} \noindent \textsc{Theorem~MK \citep{Mackenzie-Komornik2023}:}\footnote{To be precise, \cite{Mackenzie-Komornik2023} define $k$-divisibility for $k \in \{0, 1, \dots\}$ and prove this statement for $k \in \mathbb{N}$, but the same proof suffices for this stronger statement.} Fix an economy. For each $i \in N$, each $k \in [1, \infty)$, each $S \in \mathcal{S}$ that is $k$-divisible for $i$, and each $v \in [0, u_i(S)]$, if $S^* = \mathcal{G}_i(v|S)$, then
\begin{itemize}
\item $u_i(S^*) = v$, and

\item $S \setminus S^*$ is $(k-1)$-divisible for $i$.
\end{itemize}

\vspace{\baselineskip} To summarize, tails are useful for articulating divisibility in a manner that allows an individual agent to make precise cuts.

Cycles, in turn, are useful because agents with monotonic preferences agree about how to rank cycles of a given length. Moreover, if agents with monotonic preferences agree that a tail is `very' divisible, then they also agree that any superset of a maximal cycle with a `small' length is `somewhat' divisible. We formalize a slightly more general statement.

\vspace{\baselineskip} \noindent \textsc{Definition 7:} Let $S \in \mathcal{S}$ be infinite. For each nonempty $S^* \subseteq S$ and each length $\ell \in \mathbb{N}$, we say that $S^*$ is {\it $\ell$-dense in $S$} if for rank $r^* \in \mathbb{N}$ defined by $\tau_{r^*}(S) = \min S^*$, there is $r' \in \{0, 1, \dots, \ell-1\}$ such that $S^* \supseteq S_{[r^*+r'|\ell)}$. Equivalently, the requirement is that $S^*$ contains an $\ell$-cycle that is maximal with respect to set inclusion among all $\ell$-cycles in $S_{[r^*)}$.

\vspace{\baselineskip} Observe that if $S^*$ is $\ell$-dense in $S$, then each tail of $S^*$ is $\ell$-dense in $S$. Our first lemma states that if agents with monotonic preferences agree that an infinite schedule is `very' divisible, then given a `small' length $\ell$, the agents also agree that any subset of $S$ that is $\ell$-dense in $S$ is `somewhat' divisible.

\hypertarget{Lemma1}{}
\vspace{\baselineskip} \noindent \textsc{Lemma~1:} Fix an economy. For each $i \in N$ such that $u_i \in \mathcal{U}_{\text{M}}$, each $k \in [0, \infty)$, each $S \in \mathcal{S}$ that is $k$-divisible for~$i$, each $\ell \in \mathbb{N}$ such that $\ell \leq k+1$, and each $S^* \subseteq S$ that is $\ell$-dense in $S$, we have that $S^*$ is $[(\frac{1}{\ell}) \cdot (k - (\ell-1))]$-divisible for $i$.

\vspace{\baselineskip} The proof is in \hyperlink{Appendix1}{Appendix~1}. See \hyperlink{Figure1}{Figure~1} for intuition.

\hypertarget{Figure1}{}
\begin{figure*}
\centering
\captionsetup[subfigure]{justification=raggedright,singlelinecheck=false}
\subcaptionbox{Read the bottom caption first. To begin, we partition $\{t' \in S \mid t' \geq t\}$ into three $3$-cycles, which we illustrate visually using three rows.}{%
\begin{tikzpicture}[scale=0.9]
\def\gap{0.2}

\draw[draw, fill=gray!20] (0,0) rectangle (1,-1) node[pos=.5] {};
\draw[draw, fill=gray!20] (1 + \gap,-1 - \gap) rectangle (2+\gap,-2-\gap) node[pos=.5] {};
\draw[draw, fill=gray!20] (2+2*\gap,-2-2*\gap) rectangle (3+2*\gap,-3-2*\gap) node[pos=.5] {};

\draw[draw, fill=gray!20] (3+3*\gap,0) rectangle (3+3*\gap+1,-1) node[pos=.5] {};
\draw[draw, fill=gray!20] (3+3*\gap+1 + \gap,-1 - \gap) rectangle (3+3*\gap+2+\gap,-2-\gap) node[pos=.5] {};
\draw[draw, fill=gray!20] (3+3*\gap+2+2*\gap,-2-2*\gap) rectangle (3+3*\gap+3+2*\gap,-3-2*\gap) node[pos=.5] {};

\draw[draw, fill=gray!20] (6+6*\gap+0,0) rectangle (6+6*\gap+1,-1) node[pos=.5] {};
\draw[draw, fill=gray!20] (6+6*\gap+1 + \gap,-1 - \gap) rectangle (6+6*\gap+2+\gap,-2-\gap) node[pos=.5] {};
\draw[draw, fill=gray!20] (6+6*\gap+2+2*\gap,-2-2*\gap) rectangle (6+6*\gap+3+2*\gap,-3-2*\gap) node[pos=.5] {};

\draw[draw, fill=gray!20] (9+9*\gap+0,0) rectangle (9+9*\gap+1,-1) node[pos=.5] {};
\draw[draw, fill=gray!20] (9+9*\gap+1 + \gap,-1 - \gap) rectangle (9+9*\gap+2+\gap,-2-\gap) node[pos=.5] {};
\draw[draw, fill=gray!20] (9+9*\gap+2+2*\gap,-2-2*\gap) rectangle (9+9*\gap+3+2*\gap,-3-2*\gap) node[pos=.5] {};

\node at (12+12*\gap+0.5,-.5) {$\dots$};
\node at (12+12*\gap+0.5,-.5-1 -\gap) {$\dots$};
\node at (12+12*\gap+0.5,-.5-2 -2*\gap) {$\dots$};

\end{tikzpicture}%
}
\par\bigskip 
\subcaptionbox{Since $S$ is $8$-divisible, necessarily the union of the two boxes is worth at least $8 u_i(\{t\})$. Moreover, since $i$ has monotonic preferences, the left box is worth at most $2 u_i(\{t\})$; thus the right box is worth at least $6 u_i(\{t\})$.}{%
\begin{tikzpicture}[scale=0.9]
\def\gap{0.2}

\draw[draw, fill=gray!20] (0,0) rectangle (1,-1) node[pos=.5] {};
\draw[draw, fill=gray!20] (1 + \gap,-1 - \gap) rectangle (2+\gap,-2-\gap) node[pos=.5] {};
\draw[draw, fill=gray!20] (2+2*\gap,-2-2*\gap) rectangle (3+2*\gap,-3-2*\gap) node[pos=.5] {};

\draw[draw, fill=gray!20] (3+3*\gap,0) rectangle (3+3*\gap+1,-1) node[pos=.5] {};
\draw[draw, fill=gray!20] (3+3*\gap+1 + \gap,-1 - \gap) rectangle (3+3*\gap+2+\gap,-2-\gap) node[pos=.5] {};
\draw[draw, fill=gray!20] (3+3*\gap+2+2*\gap,-2-2*\gap) rectangle (3+3*\gap+3+2*\gap,-3-2*\gap) node[pos=.5] {};

\draw[draw, fill=gray!20] (6+6*\gap+0,0) rectangle (6+6*\gap+1,-1) node[pos=.5] {};
\draw[draw, fill=gray!20] (6+6*\gap+1 + \gap,-1 - \gap) rectangle (6+6*\gap+2+\gap,-2-\gap) node[pos=.5] {};
\draw[draw, fill=gray!20] (6+6*\gap+2+2*\gap,-2-2*\gap) rectangle (6+6*\gap+3+2*\gap,-3-2*\gap) node[pos=.5] {};

\draw[draw, fill=gray!20] (9+9*\gap+0,0) rectangle (9+9*\gap+1,-1) node[pos=.5] {};
\draw[draw, fill=gray!20] (9+9*\gap+1 + \gap,-1 - \gap) rectangle (9+9*\gap+2+\gap,-2-\gap) node[pos=.5] {};
\draw[draw, fill=gray!20] (9+9*\gap+2+2*\gap,-2-2*\gap) rectangle (9+9*\gap+3+2*\gap,-3-2*\gap) node[pos=.5] {};

\node at (12+12*\gap+0.5,-.5) {$\dots$};
\node at (12+12*\gap+0.5,-.5-1 -\gap) {$\dots$};
\node at (12+12*\gap+0.5,-.5-2 -2*\gap) {$\dots$};

\draw (1+.5*\gap,.5*\gap) rectangle (3+2.5*\gap,-3-2.5*\gap);
\draw (3+2.5*\gap,.5*\gap) rectangle (13+12.5*\gap,-3-2.5*\gap);

\end{tikzpicture}%
}
\par\bigskip 
\subcaptionbox{Since $i$ has monotonic preferences, the top box is worth at least $2 u_i(\{t\})$.}{%
\begin{tikzpicture}[scale=0.9]
\def\gap{0.2}

\draw[draw, fill=gray!20] (0,0) rectangle (1,-1) node[pos=.5] {};
\draw[draw, fill=gray!20] (1 + \gap,-1 - \gap) rectangle (2+\gap,-2-\gap) node[pos=.5] {};
\draw[draw, fill=gray!20] (2+2*\gap,-2-2*\gap) rectangle (3+2*\gap,-3-2*\gap) node[pos=.5] {};

\draw[draw, fill=gray!20] (3+3*\gap,0) rectangle (3+3*\gap+1,-1) node[pos=.5] {};
\draw[draw, fill=gray!20] (3+3*\gap+1 + \gap,-1 - \gap) rectangle (3+3*\gap+2+\gap,-2-\gap) node[pos=.5] {};
\draw[draw, fill=gray!20] (3+3*\gap+2+2*\gap,-2-2*\gap) rectangle (3+3*\gap+3+2*\gap,-3-2*\gap) node[pos=.5] {};

\draw[draw, fill=gray!20] (6+6*\gap+0,0) rectangle (6+6*\gap+1,-1) node[pos=.5] {};
\draw[draw, fill=gray!20] (6+6*\gap+1 + \gap,-1 - \gap) rectangle (6+6*\gap+2+\gap,-2-\gap) node[pos=.5] {};
\draw[draw, fill=gray!20] (6+6*\gap+2+2*\gap,-2-2*\gap) rectangle (6+6*\gap+3+2*\gap,-3-2*\gap) node[pos=.5] {};

\draw[draw, fill=gray!20] (9+9*\gap+0,0) rectangle (9+9*\gap+1,-1) node[pos=.5] {};
\draw[draw, fill=gray!20] (9+9*\gap+1 + \gap,-1 - \gap) rectangle (9+9*\gap+2+\gap,-2-\gap) node[pos=.5] {};
\draw[draw, fill=gray!20] (9+9*\gap+2+2*\gap,-2-2*\gap) rectangle (9+9*\gap+3+2*\gap,-3-2*\gap) node[pos=.5] {};

\node at (12+12*\gap+0.5,-.5) {$\dots$};
\node at (12+12*\gap+0.5,-.5-1 -\gap) {$\dots$};
\node at (12+12*\gap+0.5,-.5-2 -2*\gap) {$\dots$};

\draw (3+2.5*\gap,.5*\gap) rectangle (13+12.5*\gap,-1-.5*\gap);
\draw (3+2.5*\gap,.5*\gap) rectangle (13+12.5*\gap,-3-2.5*\gap);

\end{tikzpicture}%
}
\par\bigskip 
\subcaptionbox{Since $i$ has monotonic preferences, the middle box is worth at least the bottom box, which is worth at least the top box, which from the previous part is worth at least $2 u_i(\{t\})$; thus each box is worth at least $2 u_i(\{t\})$. Because $S^*$ is $3$-dense in $S$, it contains one of these boxes. Since $t \in S^*$ was arbitrary, thus $S^*$ is $2$-divisible, as desired.}{%
\begin{tikzpicture}[scale=0.9]
\def\gap{0.2}

\draw[draw, fill=gray!20] (0,0) rectangle (1,-1) node[pos=.5] {};
\draw[draw, fill=gray!20] (1 + \gap,-1 - \gap) rectangle (2+\gap,-2-\gap) node[pos=.5] {};
\draw[draw, fill=gray!20] (2+2*\gap,-2-2*\gap) rectangle (3+2*\gap,-3-2*\gap) node[pos=.5] {};

\draw[draw, fill=gray!20] (3+3*\gap,0) rectangle (3+3*\gap+1,-1) node[pos=.5] {};
\draw[draw, fill=gray!20] (3+3*\gap+1 + \gap,-1 - \gap) rectangle (3+3*\gap+2+\gap,-2-\gap) node[pos=.5] {};
\draw[draw, fill=gray!20] (3+3*\gap+2+2*\gap,-2-2*\gap) rectangle (3+3*\gap+3+2*\gap,-3-2*\gap) node[pos=.5] {};

\draw[draw, fill=gray!20] (6+6*\gap+0,0) rectangle (6+6*\gap+1,-1) node[pos=.5] {};
\draw[draw, fill=gray!20] (6+6*\gap+1 + \gap,-1 - \gap) rectangle (6+6*\gap+2+\gap,-2-\gap) node[pos=.5] {};
\draw[draw, fill=gray!20] (6+6*\gap+2+2*\gap,-2-2*\gap) rectangle (6+6*\gap+3+2*\gap,-3-2*\gap) node[pos=.5] {};

\draw[draw, fill=gray!20] (9+9*\gap+0,0) rectangle (9+9*\gap+1,-1) node[pos=.5] {};
\draw[draw, fill=gray!20] (9+9*\gap+1 + \gap,-1 - \gap) rectangle (9+9*\gap+2+\gap,-2-\gap) node[pos=.5] {};
\draw[draw, fill=gray!20] (9+9*\gap+2+2*\gap,-2-2*\gap) rectangle (9+9*\gap+3+2*\gap,-3-2*\gap) node[pos=.5] {};

\node at (12+12*\gap+0.5,-.5) {$\dots$};
\node at (12+12*\gap+0.5,-.5-1 -\gap) {$\dots$};
\node at (12+12*\gap+0.5,-.5-2 -2*\gap) {$\dots$};

\draw (3+2.5*\gap,.5*\gap) rectangle (13+12.5*\gap,-1-.5*\gap);
\draw (1+.5*\gap,-1-.5*\gap) rectangle (13+12.5*\gap,-2-1.5*\gap);
\draw (2+1.5*\gap,-2-1.5*\gap) rectangle (13+12.5*\gap,-3-2.5*\gap);

\end{tikzpicture}%
}
\par
\caption{{\it Intuition for Lemma~1.} Fix an agent $i$ with monotonic preferences, suppose that $S$ is $8$-divisible, and suppose that $S^*$ is $3$-dense in $S$. The lemma states that $S^*$ is $2$-divisible. To see this, take an arbitrary time slot $t \in S^*$. In each subfigure, we show the time slots $\{t' \in S \mid t' \geq t \}$ as shaded squares, with earlier time slots to the left. From here, follow the subcaptions.}
\end{figure*}

\hypertarget{Section3}{}
\section{Results}

Like the last section, this section is also organized into two subsections. In \hyperlink{Section3.1}{Section~3.1}, we show that there are proportional allocations if agents are sufficiently patient, while in \hyperlink{Section3.2}{Section~3.2}, we show that there are envy-free allocations if agents are sufficiently patient and moreover have weak preference for earlier time slots.

\hypertarget{Section3.1}{}
\subsection{Proportionality}

We begin by investigating the existence of proportional allocations for economies where the agents are sufficiently patient. In order to do so, we introduce \textsc{Iterative Cycle Apportionment}, a procedure where iteratively the remaining agents construct and assign a schedule.

This particular procedure belongs to a class of procedures that can each be described as follows. There are $n$ stages, and in a given stage $\mathsf{s} \in \llbracket 1, n-1 \rrbracket$, $N^\mathsf{s}$ is the set of remaining agents, $T^\mathsf{s}$ is the set of remaining time slots, and $C^\mathsf{s} \subseteq T^\mathsf{s}$ is the set of time slots under consideration. The procedure works as follows:
\begin{itemize}
\item {\it Stage $\mathsf{s}$, $\mathsf{s} \in \llbracket 1, n-1 \rrbracket$.} The agents in $N^\mathsf{s}$ start with an empty basket and consider the time slots in $C^\mathsf{s}$ in sequence. At each time slot~$t$, each remaining agent places a flag in the basket if and only if he measures the value of the basket with~$t$ to exceed~$\frac{1}{n}$. If there are no flags, then~$t$ is added to the basket and the remaining agents move to the next time slot. If there is one flag, then~$t$ is added to the basket, the basket is assigned to the agent who placed the flag, the basket's recipient exits with the basket's time slots, and we move to the next stage. If there are multiple flags, then $t$ is skipped and the remaining agents move to the next time slot. If all time slots are considered and there is no time slot with one flag, then the basket with its limit schedule---that is, the union of its schedules across all time periods---is assigned to any remaining agent whose utility for it is highest, the basket's recipient exits with the basket's time slots, and we move to the next stage.

\item {\it Stage $n$.} The remaining agent receives the remaining time slots.
\end{itemize}
As illustrated by the proof of Theorem~6 in \cite{Mackenzie-Komornik2023}, for each stage $\mathsf{s} \in \llbracket 1, n-1\rrbracket$, if the remaining agents agree that (i)~$C^\mathsf{s}$ is worth at least $\frac{1}{n}$, and (ii)~$C^\mathsf{s}$ is $1$-divisible, then the basket's recipient values the basket at least $\frac{1}{n}$ while the other remaining agents value the basket at most $\frac{1}{n}$. This observation applies to all procedures in this class.

These procedures differ in how they specify the set of time slots under consideration at each stage before the last. The simplest approach is to specify that at each such stage, {\it all} remaining time slots are under consideration, and doing so yields the \textsc{Iterative Apportionment} procedure of \cite{Mackenzie-Komornik2023}. The trouble with this particular procedure is that after the first stage, the remaining agents may not agree that what remains is $1$-divisible, in which case we cannot apply the observation from the last paragraph. Remarkably, this procedure nevertheless works in the special case that there are three agents who each have $2$-Kakeya preferences, but unfortunately the proof does not generalize.

In order to guarantee that in later stages there will be consensus that the set of time slots under consideration is $1$-divisible, we modify \textsc{Iterative Apportionment} by automatically skipping time slots to reserve them for later consideration, and doing so in a manner that still ensures there is enough value among the time slots currently under consideration. In particular, (i)~in the first stage, only the time slots from the top $(n-1)$-cycle (given $T$) are considered, (ii)~in the second stage, only the remaining time slots from the top two $(n-1)$-cycles (given $T$) are considered, and (iii)~in general, for each $\mathsf{s} \in \llbracket 1, n-1 \rrbracket$ we define $C^\mathsf{s} \equiv T^\mathsf{s} \cap [\cup_{\mathsf{s}' \leq \mathsf{s}} T_{[\mathsf{s}'|n-1)}]$. We refer to this procedure as \textsc{Iterative Cycle Apportionment}. See \hyperlink{Figure2}{Figure~2} for an example.

\hypertarget{Figure2}{}
\begin{figure*}
\centering
\captionsetup[subfigure]{justification=raggedright,singlelinecheck=false}
\subcaptionbox{First, the agents construct and assign a schedule that contains $\{1, 7\}$.}{%
\begin{tikzpicture}[scale=0.9] \footnotesize
\def\gap{0.2}

\draw[draw, fill=gray!0] (0,0) rectangle (1,-1) node[pos=.5] {$0.167$};
\draw[draw, fill=gray!20] (1 + \gap,-1 - \gap) rectangle (2+\gap,-2-\gap) node[pos=.5] {$0.139$};
\draw[draw, fill=gray!20] (2+2*\gap,-2-2*\gap) rectangle (3+2*\gap,-3-2*\gap) node[pos=.5] {$0.116$};

\draw[draw, fill=gray!20] (3+3*\gap,0) rectangle (3+3*\gap+1,-1) node[pos=.5] {$0.096$};
\draw[draw, fill=gray!20] (3+3*\gap+1 + \gap,-1 - \gap) rectangle (3+3*\gap+2+\gap,-2-\gap) node[pos=.5] {$0.080$};
\draw[draw, fill=gray!20] (3+3*\gap+2+2*\gap,-2-2*\gap) rectangle (3+3*\gap+3+2*\gap,-3-2*\gap) node[pos=.5] {$0.067$};

\draw[draw, fill=gray!0] (6+6*\gap+0,0) rectangle (6+6*\gap+1,-1) node[pos=.5] {$0.056$};
\draw[draw, fill=gray!20] (6+6*\gap+1 + \gap,-1 - \gap) rectangle (6+6*\gap+2+\gap,-2-\gap) node[pos=.5] {$0.047$};
\draw[draw, fill=gray!20] (6+6*\gap+2+2*\gap,-2-2*\gap) rectangle (6+6*\gap+3+2*\gap,-3-2*\gap) node[pos=.5] {$0.039$};

\draw[draw, fill=gray!20] (9+9*\gap+0,0) rectangle (9+9*\gap+1,-1) node[pos=.5] {$0.032$};
\draw[draw, fill=gray!20] (9+9*\gap+1 + \gap,-1 - \gap) rectangle (9+9*\gap+2+\gap,-2-\gap) node[pos=.5] {$0.027$};
\draw[draw, fill=gray!20] (9+9*\gap+2+2*\gap,-2-2*\gap) rectangle (9+9*\gap+3+2*\gap,-3-2*\gap) node[pos=.5] {$0.022$};

\node at (12+12*\gap+0.5,-.5) {$\dots$};
\node at (12+12*\gap+0.5,-.5-1 -\gap) {$\dots$};
\node at (12+12*\gap+0.5,-.5-2 -2*\gap) {$\dots$};

\draw[dashed] (-.5*\gap,.5*\gap) -- (13+12.5*\gap,.5*\gap);
\draw[] (-.5*\gap,-1-.5*\gap) -- (13+12.5*\gap,-1-.5*\gap);

\end{tikzpicture}%
}
\par\medskip 
\subcaptionbox{Second, the agents construct and assign a schedule that contains $\{2, 4\}$.}{%
\begin{tikzpicture}[scale=0.9] \footnotesize
\def\gap{0.2}

\draw[draw, fill=gray!0] (0,0) rectangle (1,-1) node[pos=.5] {};
\draw[draw, fill=gray!0] (1 + \gap,-1 - \gap) rectangle (2+\gap,-2-\gap) node[pos=.5] {$0.139$};
\draw[draw, fill=gray!20] (2+2*\gap,-2-2*\gap) rectangle (3+2*\gap,-3-2*\gap) node[pos=.5] {$0.116$};

\draw[draw, fill=gray!0] (3+3*\gap,0) rectangle (3+3*\gap+1,-1) node[pos=.5] {$0.096$};
\draw[draw, fill=gray!20] (3+3*\gap+1 + \gap,-1 - \gap) rectangle (3+3*\gap+2+\gap,-2-\gap) node[pos=.5] {$0.080$};
\draw[draw, fill=gray!20] (3+3*\gap+2+2*\gap,-2-2*\gap) rectangle (3+3*\gap+3+2*\gap,-3-2*\gap) node[pos=.5] {$0.067$};

\draw[draw, fill=gray!0] (6+6*\gap+0,0) rectangle (6+6*\gap+1,-1) node[pos=.5] {};
\draw[draw, fill=gray!20] (6+6*\gap+1 + \gap,-1 - \gap) rectangle (6+6*\gap+2+\gap,-2-\gap) node[pos=.5] {$0.047$};
\draw[draw, fill=gray!20] (6+6*\gap+2+2*\gap,-2-2*\gap) rectangle (6+6*\gap+3+2*\gap,-3-2*\gap) node[pos=.5] {$0.039$};

\draw[draw, fill=gray!20] (9+9*\gap+0,0) rectangle (9+9*\gap+1,-1) node[pos=.5] {$0.032$};
\draw[draw, fill=gray!20] (9+9*\gap+1 + \gap,-1 - \gap) rectangle (9+9*\gap+2+\gap,-2-\gap) node[pos=.5] {$0.027$};
\draw[draw, fill=gray!20] (9+9*\gap+2+2*\gap,-2-2*\gap) rectangle (9+9*\gap+3+2*\gap,-3-2*\gap) node[pos=.5] {$0.022$};

\node at (12+12*\gap+0.5,-.5) {$\dots$};
\node at (12+12*\gap+0.5,-.5-1 -\gap) {$\dots$};
\node at (12+12*\gap+0.5,-.5-2 -2*\gap) {$\dots$};

\draw[dashed] (-.5*\gap,-1-.5*\gap) -- (13+12.5*\gap,-1-.5*\gap);
\draw[] (-.5*\gap,-2-1.5*\gap) -- (13+12.5*\gap,-2-1.5*\gap);

\end{tikzpicture}%
}
\par\medskip 
\subcaptionbox{Third, the agents construct and assign a schedule that contains $\{3, 5, 8\}$.}{%
\begin{tikzpicture}[scale=0.9] \footnotesize
\def\gap{0.2}

\draw[draw, fill=gray!0] (0,0) rectangle (1,-1) node[pos=.5] {};
\draw[draw, fill=gray!0] (1 + \gap,-1 - \gap) rectangle (2+\gap,-2-\gap) node[pos=.5] {};
\draw[draw, fill=gray!0] (2+2*\gap,-2-2*\gap) rectangle (3+2*\gap,-3-2*\gap) node[pos=.5] {$0.116$};

\draw[draw, fill=gray!0] (3+3*\gap,0) rectangle (3+3*\gap+1,-1) node[pos=.5] {};
\draw[draw, fill=gray!0] (3+3*\gap+1 + \gap,-1 - \gap) rectangle (3+3*\gap+2+\gap,-2-\gap) node[pos=.5] {$0.080$};
\draw[draw, fill=gray!20] (3+3*\gap+2+2*\gap,-2-2*\gap) rectangle (3+3*\gap+3+2*\gap,-3-2*\gap) node[pos=.5] {$0.067$};

\draw[draw, fill=gray!0] (6+6*\gap+0,0) rectangle (6+6*\gap+1,-1) node[pos=.5] {};
\draw[draw, fill=gray!0] (6+6*\gap+1 + \gap,-1 - \gap) rectangle (6+6*\gap+2+\gap,-2-\gap) node[pos=.5] {$0.047$};
\draw[draw, fill=gray!20] (6+6*\gap+2+2*\gap,-2-2*\gap) rectangle (6+6*\gap+3+2*\gap,-3-2*\gap) node[pos=.5] {$0.039$};

\draw[draw, fill=gray!20] (9+9*\gap+0,0) rectangle (9+9*\gap+1,-1) node[pos=.5] {$0.032$};
\draw[draw, fill=gray!20] (9+9*\gap+1 + \gap,-1 - \gap) rectangle (9+9*\gap+2+\gap,-2-\gap) node[pos=.5] {$0.027$};
\draw[draw, fill=gray!20] (9+9*\gap+2+2*\gap,-2-2*\gap) rectangle (9+9*\gap+3+2*\gap,-3-2*\gap) node[pos=.5] {$0.022$};

\node at (12+12*\gap+0.5,-.5) {$\dots$};
\node at (12+12*\gap+0.5,-.5-1 -\gap) {$\dots$};
\node at (12+12*\gap+0.5,-.5-2 -2*\gap) {$\dots$};

\draw[dashed] (-.5*\gap,-2-1.5*\gap) -- (13+12.5*\gap,-2-1.5*\gap);
\draw[] (-.5*\gap,-3-2.5*\gap) -- (13+12.5*\gap,-3-2.5*\gap);

\end{tikzpicture}%
}
\par\medskip 
\subcaptionbox{Finally, the remaining time slots are assigned to the remaining agent.}{%
\begin{tikzpicture}[scale=0.9] \footnotesize
\def\gap{0.2}

\draw[draw, fill=gray!0] (0,0) rectangle (1,-1) node[pos=.5] {};
\draw[draw, fill=gray!0] (1 + \gap,-1 - \gap) rectangle (2+\gap,-2-\gap) node[pos=.5] {};
\draw[draw, fill=gray!0] (2+2*\gap,-2-2*\gap) rectangle (3+2*\gap,-3-2*\gap) node[pos=.5] {};

\draw[draw, fill=gray!0] (3+3*\gap,0) rectangle (3+3*\gap+1,-1) node[pos=.5] {};
\draw[draw, fill=gray!0] (3+3*\gap+1 + \gap,-1 - \gap) rectangle (3+3*\gap+2+\gap,-2-\gap) node[pos=.5] {};
\draw[draw, fill=gray!0] (3+3*\gap+2+2*\gap,-2-2*\gap) rectangle (3+3*\gap+3+2*\gap,-3-2*\gap) node[pos=.5] {$0.067$};

\draw[draw, fill=gray!0] (6+6*\gap+0,0) rectangle (6+6*\gap+1,-1) node[pos=.5] {};
\draw[draw, fill=gray!0] (6+6*\gap+1 + \gap,-1 - \gap) rectangle (6+6*\gap+2+\gap,-2-\gap) node[pos=.5] {};
\draw[draw, fill=gray!0] (6+6*\gap+2+2*\gap,-2-2*\gap) rectangle (6+6*\gap+3+2*\gap,-3-2*\gap) node[pos=.5] {$0.039$};

\draw[draw, fill=gray!0] (9+9*\gap+0,0) rectangle (9+9*\gap+1,-1) node[pos=.5] {$0.032$};
\draw[draw, fill=gray!0] (9+9*\gap+1 + \gap,-1 - \gap) rectangle (9+9*\gap+2+\gap,-2-\gap) node[pos=.5] {$0.027$};
\draw[draw, fill=gray!0] (9+9*\gap+2+2*\gap,-2-2*\gap) rectangle (9+9*\gap+3+2*\gap,-3-2*\gap) node[pos=.5] {$0.022$};

\node at (12+12*\gap+0.5,-.5) {$\dots$};
\node at (12+12*\gap+0.5,-.5-1 -\gap) {$\dots$};
\node at (12+12*\gap+0.5,-.5-2 -2*\gap) {$\dots$};

\draw[dashed] (-.5*\gap,-3-2.5*\gap) -- (13+12.5*\gap,-3-2.5*\gap);

\end{tikzpicture}%
}
\par
\caption{{\it Example of Iterative Cycle Apportionment.} In this example there are four agents, each with $5$-Kakeya preferences. The time slots $\{1, 2, \dots\}$ are illustrated as squares, with earlier time slots to the left, and they are partitioned into three $3$-cycles. Each agent has a geometric utility function with a discount factor that is at least $\frac{5}{6}$ and extremely close to $\frac{5}{6}$: for each time slot in the figure and each agent, the number on the time slot's square is the value that the agent assigns to it, rounded to the nearest thousandth. The only purpose of our assumption that the discount factors are extremely close to each other is the simplification of the figure, as it allows us to write one rounded number on each square that is used by all agents, but in general the procedure only requires the agents to be sufficiently patient. At each stage, previously considered time slots are above the dotted line, reserved time slots are below the solid line, blank squares are removed in earlier stages, white squares with numbers are removed in the current stage, and shaded squares remain after the current stage.}
\end{figure*}

In order to guarantee \textsc{Iterative Cycle Apportionment} constructs a proportional allocation, we require that (i)~each agent deems each $(n-1)$-cycle of $T$ to be $1$-divisible, and (ii)~each agent has monotonic preferences. Moreover, for an arbitrary economy we can construct an associated economy with monotonic preferences, and if this associated economy satisfies {\it both} requirements, then we can construct a proportional allocation for the original economy by applying \textsc{Iterative Cycle Apportionment} to the associated economy and modifying the resulting allocation.

\vspace{\baselineskip} \noindent \textsc{Definition 8:} Fix an economy and a utility function $u_0$. The \emph{monotonic reordering of~$u_0$} is the monotonic utility function $u_0^{\text{M}} \in \mathcal{U}_{\text{M}}$ formed by arranging the values of $u_0$ in non-increasing order. Formally, we have (i)~$u_0^{\text{M}}(\{1\}) \geq u_0^{\text{M}}(\{2\}) \geq \dots$, and (ii)~for each $v \in (0, 1]$, $|\{t\in T\mid u_0(\{t\}) = v\}| = |\{t\in T\mid u_0^{\text{M}}(\{t\}) = v\}|$, where for emphasis this cardinality must be finite as $v$ is positive and $\sum_{t \in T} u_0(\{t\}) = 1$. We refer to $(n, u^{\text{M}}) = (n, (u^{\text{M}}_i)_{i \in N})$ as the {\it monotonic economy associated with $(n, u)$}.

\vspace{\baselineskip} First, we show that patience is preserved under monotonic reordering.

\hypertarget{Lemma2}{}
\vspace{\baselineskip} \noindent \textsc{Lemma~2:} 
For each $k \in [0,\infty)$ and each $u_0 \in \mathcal{U}_k$, $u_0^{\text{M}} \in\mathcal{U}_k \cap \mathcal{U}_{\text{M}}$.

\vspace{\baselineskip} The proof is in \hyperlink{Appendix2}{Appendix~2}. Second, we show that a proportional allocation $\pi^{\text{M}}$ for the associated economy with monotonic preferences can be used to construct a proportional allocation~$\pi$ for the original economy. In particular, in the original economy, we consider the time slots in sequence, and at each time slot $t$ we ask the recipient of $t$ according to~$\pi^{\text{M}}$ to select his favorite of the remaining time slots; we then assign any time slots that are not selected to the first agent. Since each agent's utility for $\pi$ in the original economy is at least as high as his utility for $\pi^{\text{M}}$ in the associated economy with monotonic preferences, $\pi$ is proportional.

\hypertarget{Lemma3}{}
\vspace{\baselineskip} \noindent \textsc{Lemma~3:} For each economy $(n, u)$, if $(n, u^{\text{M}})$ has a proportional allocation, then $(n, u)$ does as well.

\vspace{\baselineskip} The proof is in \hyperlink{Appendix2}{Appendix~2}. Using (i)~the preceding two lemmas, (ii)~the implication of \hyperlink{Lemma1}{Lemma~1} that each agent with monotonic $(2n-3)$-Kakeya preferences deems each $(n-1)$-cycle to be $1$-divisible, and (iii)~an argument establishing that \textsc{Iterative Cycle Apportionment} constructs a proportional allocation under the claimed requirements, we establish our first main result.\footnote{We remark that we could have also established this theorem using a simpler but longer variant of \textsc{Iterative Cycle Apportionment}: whenever there is exactly one flag, skip the current time slot instead of taking it and assigning the basket. Indeed, this merges two cases into one at the cost of condemning all remaining agents to consider all time slots under consideration at each stage. Interestingly, while this modification to \textsc{Iterative Cycle Apportionment} does not impact our argument, the same modification to \textsc{Iterative Apportionment} does break the three-agent argument of \cite{Mackenzie-Komornik2023}: there it is crucial that at a time slot with multiple flags in the second stage, at least one of the agents who placed a flag also did so at the same time slot in the first stage, but with the modification it is possible that a single agent placed a flag at this time slot in the first stage and then left with the first stage's basket.}

\hypertarget{Theorem1}{}
\vspace{\baselineskip} \noindent \textsc{Theorem~1:} Fix an economy with at least two agents. If for each $i \in N$ we have $u_i \in \mathcal{U}_{2n-3}$, then there is a proportional allocation.

\vspace{\baselineskip} The proof is in \hyperlink{Appendix2}{Appendix~2}. As remarked in \hyperlink{Section1}{Section~1}, in order for knowledge that all utility functions belong to $\mathcal{U}_k$ to imply there is a proportional allocation, it is necessary that $k \geq n-1$; by \hyperlink{Theorem1}{Theorem~1}, it is sufficient that $k \geq 2n-3$. Though the lower bound and upper bound on the level of patience that guarantees proportional allocations do not coincide, they are both linear in the number of agents, so the level of patience that guarantees proportional allocations is linear in the number of agents, and moreover our bounds coincide when $n=2$. If moreover it is known that all utility functions are geometric discounting, then using the remark from \hyperlink{Section1}{Section~1}, we can translate these statements into the language of discount factors: in order for knowledge that each discount factor is at least $\delta^*$ to imply there is a proportional allocation, it is necessary that $\delta^* \geq \frac{n-1}{n}$ and it is sufficient that $\delta^* \geq \frac{2n-3}{2n-2}$.

By Observation~1 of \cite{Mackenzie-Komornik2023}, if there is a proportional allocation, then there is a proportional allocation that is efficient; we thus immediately have the following.

\hypertarget{Corollary1}{}
\vspace{\baselineskip} \noindent \textsc{Corollary~1:} Fix an economy with at least two agents. If for each $i \in N$ we have $u_i \in \mathcal{U}_{2n-3}$, then there is a proportional allocation that is efficient.

\hypertarget{Section3.2}{}
\subsection{No envy}

Unfortunately, \textsc{Iterative Cycle Apportionment} need not construct an envy-free allocation: an agent who receives his schedule in an earlier stage may envy an agent who receives his schedule in a later stage. In order to guarantee that there is an envy-free allocation, we require the agents to be so patient that they can construct such an allocation using a protocol for fair division of a perfectly divisible cake.

In particular, let us say that a {\it cake (query) protocol}, $\mathsf{Q}$, is a method for a mediator who initially does not know the preferences of the agents to iteratively query the agents and then select an allocation of the classic interval cake $(0, 1]$. The details of precisely which cake protocols we allow are not essential to the main ideas in this paper, but we include a summary of the key details at this point for the interested reader; see \hyperlink{Appendix4}{Appendix~4} for all formal definitions.

\vspace{\baselineskip} \noindent \textsc{Cake protocol summary:} In this summary, we explain the queries that we allow, the cellular restriction we impose, what we require a protocol to accomplish in order to be declared envy-free, and how we measure costs. To avoid confusion, we refer to preference representations in the context of cakes as {\it cake measures}, which are measures on $(0, 1]$ with some additional structure described below, and we write $\mu_i$ for the cake measure of agent $i$.

First, in the classic Robertson-Webb model \citep{Robertson-Webb1998}, there are two kinds of queries: (i)~in an \textsc{Evaluate} query, an agent $i$ is given a sub-interval $(x, y] \subseteq (0, 1]$ and asked to evaluate its worth according to $\mu_i$, and (ii)~in a \textsc{Cut} query, an agent $i$ is given a point $x \in [0,1)$ and a value $v \in [0,1]$, then asked to cut at a point $y \in [x, 1]$ such that $\mu_i((x, y]) = v$ (if possible). Many protocols in the literature use variants of these queries, and these variants can be translated from one to another via small arguments; see \cite{Cseh-Fleiner2020} for a nice discussion of this point. We only use one kind of query, a cut followed by automatic evaluations by everybody, with a superficial change that does not impact our analysis.\footnote{The superficial change is that we work with target proportions instead of target values, simply for consistency with the more general formal definition of protocol that we adapt for our purposes here \citep{Brandl-Mackenzie2026}, to facilitate comparison with that paper. This is superficial because our mediator always knows an appraised partition and asks an agent to cut a cell of that partition; see the discussion of our cellular requirement in the next paragraph.}

Second, we require the mediator to always ask the cutter to cut an interval that is a cell of the partition given by the current cut points, and call a protocol that satisfies this requirement {\it cellular}. The cake protocols that we use from the literature all work exclusively with $0$, $1$, and previous cut points, and therefore satisfy this restriction after a superficial change; thus this restriction allows us to simplify our arguments at no cost to our main message. As a result, at each stage of a cake protocol, the information available to the mediator---that is, the full history of queries and responses---determines an {\it appraised partition}: the partition given by the cuts thus far, together with an appraisal by each agent of each cell's value.

Third, we are generous in declaring a cellular cake protocol to be {\it envy-free}: it need only produce envy-free allocations for preference profiles given by positive piecewise-uniform measures. This strengthens our result: as long as a cake protocol from the literature is known to accomplish this much, we can use it.

Finally, we measure costs of our cake protocols using our queries. Equivalently, we measure costs of cellular protocols that use Robertson-Webb queries using cut queries. More precisely, we let $\mathsf{cost}(\mathsf{Q})$ denote the worst-case number of queries used by $\mathsf{Q}$, and we are particularly interested in envy-free cellular protocols that are {\it bounded}: there is a guarantee on the number of required cuts that holds across all measure profiles, so that $\mathsf{cost}(\mathsf{Q})$ is finite.

\vspace{\baselineskip} At a high level, our goal is to {\it import} a cake protocol to our setting. Indeed, given an input utility function profile $u$, the rough idea is that we begin with the coarsest appraised partitions of the cake and the set of all time slots, match their cells, and then iteratively (i)~use $\mathsf{Q}$ to select a cake query, (ii)~use the matching to execute an analogous cut {\it on the time slots}, (iii)~use $u$ to appraise the new time partition, and (iv)~automatically cut the given cake cell at its midpoint, match the new cake cells to the new time cells, and update the cake appraisal accordingly. This is done until $\mathsf{Q}$ selects a cake allocation, at which point (after cleaning up the cake allocation to respect the partition if necessary) we select the matching time allocation. We remark that under the interpretation that the given $u$ is {\it unknown}, we can interpret the imported cake protocol as a ``protocol" for the time problem, though for brevity we do not formalize this by providing a variant of \hyperlink{Appendix4}{Appendix~4} dedicated to protocols for the time problem.

The challenge for importing a cake protocol to our setting is that for perfectly divisible cakes, any requested cut of a set $S$ is successful so long as the target value belongs to $[0, \mu_i(S)]$, but in our setting that need not be the case. Moreover, even if $i$ is given a set $S$ that he deems $k$-divisible for $k \geq 1$, the \textsc{Greedy Algorithm} does not suffice for our purposes: though $i$ can cut a set $S^*$ that is worth the target value while guaranteeing that he deems $S \setminus S^*$ to be $(k-1)$-divisible, there is no guarantee that another agent deems $S \setminus S^*$ to be divisible at all, and there is no guarantee that any agent deems $S^*$ to be divisible at all.

To solve this problem, we introduce an alternative to the \textsc{Greedy Algorithm}. Intuitively, when cutting $S^*$ from a schedule $S$ that all agree is $k$-divisible, the \textsc{Greedy Algorithm} can be used as long as $k \geq 1$, and promises (i)~$S \setminus S^*$ is $(k-1)$-divisible for the cutter, (ii)~nothing about the divisibility of $S \setminus S^*$ for other agents, and (iii)~nothing about the divisibility of $S^*$ for anybody. Our alternative, which we refer to as the \textsc{Tripartition Algorithm}, requires monotonic preferences and $k \geq 5$, and promises (i)~$S \setminus S^*$ is $\frac{k-2}{3}$-divisible for everybody, and (ii)~$S^*$ is $\frac{k-2}{3}$-divisible for everybody. We give the formal definition below; see \hyperlink{Figure3}{Figure~3} for an example.

\vspace{\baselineskip} \noindent \textsc{Definition 9:} {\it Tripartition Algorithm.} Fix an economy. For each $i \in N$, each infinite $S \in \mathcal{S}$, and each $v \in (0, u_i(S))$, let $\mathcal{T}_i(v|S) \subseteq S$ denote the {\it tripartition schedule (for agent~$i$ given source $S$ and target $v$)} constructed in two cases as follows. First, if $v \leq (\frac{1}{2}) \cdot u_i(S)$, then do the following.
\begin{itemize}
\item Define $r \equiv \min \{r' \in \mathbb{N} \mid (\frac{1}{3}) \cdot u_i(S_{[r')}) \leq v \}$. Observe that this is well-defined because $v > 0$ and $u_i(S) \leq 1$.

\item Define $S^{\text{sort}} \equiv S_{[r|3)}$, $S^{\text{skip}} \equiv S_{[r+1|3)}$, and $S^{\text{take}} \equiv S_{[r+2|3)}$. Observe that these three schedules partition $S_{[r)}$.

\item We define $\mathcal{T}_i(v|S) \subseteq S$ in three subcases: (i)~if $v - u_i(S^{\text{take}}) < 0$, then $\mathcal{T}_i(v|S) \equiv S^{\text{take}}$, (ii)~if $v - u_i(S^{\text{take}}) > u_i(S^{\text{sort}})$, then $\mathcal{T}_i(v|S) \equiv S^{\text{take}} \cup S^{\text{sort}}$, and (iii)~if $v - u_i(S^{\text{take}}) \in [0, u_i(S^{\text{sort}})]$, then $\mathcal{T}_i(v|S) \equiv S^{\text{take}} \cup \mathcal{G}_i(v - u_i(S^{\text{take}}) | S^{\text{sort}})$. We will prove that if $u_i \in \mathcal{U}_{\text{M}}$ and if $S$ is $5$-divisible for $i$, then the third subcase applies: the target value for the \textsc{Greedy Algorithm} is not too low or too high.
\end{itemize}
Second, if $v > (\frac{1}{2}) \cdot u_i(S)$, then define $\mathcal{T}_i(v|S) \equiv S \setminus [\mathcal{T}_i(u_i(S) - v|S)]$.

\hypertarget{Proposition1}{}
\vspace{\baselineskip} \noindent \textsc{Proposition~1:} Fix an economy. For each pair $i, j \in N$ such that $u_i \in \mathcal{U}_{\text{M}}$ and $u_j \in \mathcal{U}_{\text{M}}$, each $k \in [5, \infty)$, each $S \in \mathcal{S}$ that is $k$-divisible for both $i$ and $j$, and each $v \in (0, u_i(S))$, if $S^* = \mathcal{T}_i(v|S)$, then
\begin{itemize}
\item $u_i(S^*) = v$,

\item $S^*$ is $[(\frac{1}{3}) \cdot (k - 2)]$-divisible for both $i$ and $j$, and

\item $S \setminus S^*$ is $[(\frac{1}{3}) \cdot (k - 2)]$-divisible for both $i$ and $j$.
\end{itemize}

\vspace{\baselineskip} The proof is in \hyperlink{Appendix3}{Appendix~3}. Given a bounded envy-free cellular cake protocol $\mathsf{Q}$, it follows that if the agents are sufficiently patient, then we can import $\mathsf{Q}$ to our setting by (i)~making each cut before the final step with the \textsc{Tripartition Algorithm}, and (ii)~using the \textsc{Greedy Algorithm} for the final step. Indeed, if the agents are sufficiently patient, then (i)~before each step of the protocol except possibly the final one, the agents agree that each member of the partition is at least $5$-divisible, and (ii)~before the final step of the protocol, the agents agree that each member of the partition is $1$-divisible. We remark that in this sense, the \textsc{Tripartition Algorithm} provides a useful new method of cutting for a single agent, or {\it knife}, highlighting that this discrete problem of fairly taking turns differs from the continuous problem of interval cake division in that here we are not initially given all the knives we might ever need for construction, but rather here we must construct even the knives.

\hypertarget{Figure3}{}
\begin{figure*}
\centering
\begin{tikzpicture}[scale=0.9] \footnotesize
\def\gap{0.2}

\draw[draw, fill=gray!20] (0,0) rectangle (1,-1) node[pos=.5] {$0.167$};
\draw[draw, fill=gray!20] (1 + \gap,-1 - \gap) rectangle (2+\gap,-2-\gap) node[pos=.5] {$0.139$};
\draw[draw, fill=gray!20] (2+2*\gap,-2-2*\gap) rectangle (3+2*\gap,-3-2*\gap) node[pos=.5] {$0.116$};

\draw[draw, fill=gray!20] (3+3*\gap,0) rectangle (3+3*\gap+1,-1) node[pos=.5] {$0.096$};
\draw[draw, fill=gray!20] (3+3*\gap+1 + \gap,-1 - \gap) rectangle (3+3*\gap+2+\gap,-2-\gap) node[pos=.5] {$0.080$};
\draw[draw, fill=gray!00] (3+3*\gap+2+2*\gap,-2-2*\gap) rectangle (3+3*\gap+3+2*\gap,-3-2*\gap) node[pos=.5] {$0.067$};

\draw[draw, fill=gray!0] (6+6*\gap+0,0) rectangle (6+6*\gap+1,-1) node[pos=.5] {$0.056$};
\draw[draw, fill=gray!20] (6+6*\gap+1 + \gap,-1 - \gap) rectangle (6+6*\gap+2+\gap,-2-\gap) node[pos=.5] {$0.047$};
\draw[draw, fill=gray!0] (6+6*\gap+2+2*\gap,-2-2*\gap) rectangle (6+6*\gap+3+2*\gap,-3-2*\gap) node[pos=.5] {$0.039$};

\draw[draw, fill=gray!20] (9+9*\gap+0,0) rectangle (9+9*\gap+1,-1) node[pos=.5] {$0.032$};
\draw[draw, fill=gray!20] (9+9*\gap+1 + \gap,-1 - \gap) rectangle (9+9*\gap+2+\gap,-2-\gap) node[pos=.5] {$0.027$};
\draw[draw, fill=gray!0] (9+9*\gap+2+2*\gap,-2-2*\gap) rectangle (9+9*\gap+3+2*\gap,-3-2*\gap) node[pos=.5] {$0.022$};

\node at (12+12*\gap+0.5,-.5) {$\dots$};
\node at (12+12*\gap+0.5,-.5-1 -\gap) {$\dots$};
\node at (12+12*\gap+0.5,-.5-2 -2*\gap) {$\dots$};

\draw (3+2.5*\gap,.5*\gap) rectangle (13+12.5*\gap,-1-.5*\gap);
\draw (3+2.5*\gap,-1-.5*\gap) rectangle (13+12.5*\gap,-2-1.5*\gap);
\draw (3+2.5*\gap,-2-1.5*\gap) rectangle (13+12.5*\gap,-3-2.5*\gap);

\node at (13+13*\gap+0.5,-.5) {sort};
\node at (13+13*\gap+0.5,-.5-1 -\gap) {skip};
\node at (13+13*\gap+0.5,-.5-2 -2*\gap) {take};

\end{tikzpicture}
\caption{{\it Example of Tripartition Algorithm.} In this example, agent $i$ has $5$-Kakeya preferences. The time slots $\{1, 2, \dots\}$ are illustrated as squares, with earlier time slots to the left, and they are partitioned into three $3$-cycles. Agent $i$ has a geometric utility function with discount factor $\frac{5}{6}$: for each time slot in the figure, the number on the time slot's square is the value that $i$ assigns to it, rounded to the nearest thousandth. The figure shows how $i$ constructs a schedule worth $0.22$ from the set of all time slots using the \textsc{Tripartition Algorithm}. In particular, the procedure calculates the rank $r=4$, then uses that rank to partition the time slots $\{4, 5, \dots\}$ into the schedules $S^{\text{sort}}$, $S^{\text{skip}}$, and $S^{\text{take}}$; these are the top box, the middle box, and the bottom box, respectively. To complete the procedure, agent $i$ takes all of $S^{\text{take}}$, then takes the remaining value he requires from $S^{\text{sort}}$ using the \textsc{Greedy Algorithm} while leaving all of $S^{\text{skip}}$; white squares are removed while shaded squares remain.}
\end{figure*}

\vspace{\baselineskip} \noindent \textsc{Definition 10:} {\it Divisibility requirement.} Define the divisibility requirement $\mathsf{d}$ such that for each number of cuts $c \in \mathbb{N}$, $\mathsf{d}(c) \equiv 2 \cdot 3^{c-1}-1$. Intuitively, this is the amount of divisibility required on the set of all time slots in order to successfully import an envy-free cellular cake protocol that uses at most $c$ cuts.\footnote{Indeed, following our remarks about {\it importing}, we use the \textsc{Tripartition~Algorithm} for every cut but the final one and use the \textsc{Greedy~Algorithm} for the final cut. By \hyperlink{Proposition1}{Proposition~1}, for each $k \in [5, \infty)$, the \textsc{Tripartition~Algorithm} divides a schedule all deem $k$-divisible into two parts that all deem $\frac{k-2}{3}$-divisible, and we require the final schedule that is cut to be $1$-divisible in order to apply \hyperlink{TheoremMK}{Theorem~MK}. The divisibility requirement promises that we can make $c$ such cuts successfully if initially all agree that $T$ is $\mathsf{d}(c)$-divisible: (i)~$\mathsf{d}(1)=1$, and (ii)~for each $c \in \mathbb{N}$ such that $c>1$, $\mathsf{d}(c) = 3 \cdot \mathsf{d}(c-1)+2 \geq 5$.}

\vspace{\baselineskip} Using this notion of divisibility requirement, as well as the notions of {\it associated cake setting}, {\it envy-free cellular cake protocol}, and {\it cost of a cake protocol} from \hyperlink{Appendix4}{Appendix~4}, we can state our theorem for fairly taking turns in an envy-free manner.

\hypertarget{Theorem2}{}
\vspace{\baselineskip} \noindent \textsc{Theorem~2:} Fix an economy. For each number of cuts $c \in \mathbb{N}$, if (i)~the associated cake setting has an envy-free cellular cake protocol $\mathsf{Q}$ such that $\mathsf{cost}(\mathsf{Q}) \leq c$, and (ii)~for each $i \in N$ we have $u_i \in \mathcal{U}_{\mathsf{d}(c)} \cap \mathcal{U}_{\text{M}}$, then there is an envy-free allocation.

\vspace{\baselineskip} The proof is in \hyperlink{Appendix3}{Appendix~3}. We conclude this section by applying \hyperlink{Theorem2}{Theorem~2} to (i)~the ancient notion of \textsc{Divide and Choose} for $n=2$, (ii)~the \textsc{Selfridge-Conway} protocol for $n=3$, and (iii)~the \textsc{Psomas} protocol for arbitrary $n$ \citep{Psomas2026}. For the general case, we also briefly discuss the \textsc{Aziz-Mackenzie} protocol \citep{Aziz-Mackenzie2016}, the \textsc{Sokolov} protocol \citep{Sokolov2023}, and the \textsc{Ye-Bai} protocol \citep{Ye-Bai2026}. We begin with $n=2$.

\hypertarget{Corollary2}{}
\vspace{\baselineskip} \noindent \textsc{Corollary~2:} Fix an economy with $n=2$. If for each $i \in N$ we have $u_i \in \mathcal{U}_1 \cap \mathcal{U}_{\text{M}}$, then there is an envy-free allocation.

\vspace{\baselineskip} Indeed, according to our formal definitions, \textsc{Divide and Choose} is an envy-free cellular cake protocol whose cost is one: the mediator asks one agent to divide the entire cake into two equally valuable parts, then both parts are automatically evaluated by the other agent, and finally the mediator is able to select an envy-free allocation by giving the other agent one of his favorite parts and giving the first agent the other part. The corollary then follows from \hyperlink{Theorem2}{Theorem~2} because $\mathsf{d}(1) = 1$. In fact, \hyperlink{Corollary2}{Corollary~2} is also implied by Corollary~1 of \cite{Mackenzie-Komornik2023}, because \textsc{Divide and Choose} works even if neither agent has monotonic preferences and only the first agent has preferences in $\mathcal{U}_1$: we only need one agent to use the \textsc{Greedy Algorithm}.

We now turn to the case of three agents.

\hypertarget{Corollary3}{}
\vspace{\baselineskip} \noindent \textsc{Corollary~3:} Fix an economy with $n=3$. If for each $i \in N$ we have $u_i \in \mathcal{U}_{161} \cap \mathcal{U}_{\text{M}}$, then there is an envy-free allocation.

\vspace{\baselineskip} Indeed, according to our formal definitions, the \textsc{Selfridge-Conway} protocol is an envy-free cellular cake protocol whose cost is five. A concise statement of this remarkable protocol appears in \cite{Branzei-Caragiannis-Kurokawa-Procaccia2016}, and we adapt their statement in order to emphasize our formal cake protocol requirements: (i)~agent~1 cuts the cake with target proportion $\frac{1}{3}$, then cuts the remainder with target proportion $\frac{1}{2}$, to construct three {\it pieces} he deems equivalent; (ii)~if agent~2 has a unique favorite piece, then he cuts his favorite piece with the target proportion equal to one minus the ratio of his second highest piece value to his highest piece value, so that he is indifferent between his two favorite pieces after the produced {\it trimmings} are removed; (iii)~agent~3 takes his favorite piece; (iv)~agent~2 takes the trimmed piece if he did trim a piece and that piece remains available, and otherwise takes his favorite remaining piece; (v)~agent~1 takes the remaining piece, and at this point we are done unless agent~2 produced trimmings, in which case (vi)~let $i \in \{2, 3\}$ denote the agent who received the trimmed piece, and let $j$ denote the other member of $\{2, 3\}$; (vii)~agent~$j$ cuts the trimmings with target proportion $\frac{1}{3}$, then cuts the remainder with target proportion $\frac{1}{2}$, to construct three {\it pieces} he deems equivalent; (viii)~agent $i$ takes his favorite piece; (ix)~agent~1 takes his favorite remaining piece; and (x)~agent $j$ takes the final piece. This indeed involves at most five cuts, and the corollary then follows from \hyperlink{Theorem2}{Theorem~2} because $\mathsf{d}(5) = 161$. We illustrate how it works imported to our setting in \hyperlink{Figure4}{Figure~4}, and argue that all details in this figure apply for certain economies with geometric utility functions below.

\hypertarget{Figure4}{}
\begin{figure*}
\centering
\begin{tikzpicture}[scale=0.9] \footnotesize

\newcommand{\SCstep}[1]{\tikz[baseline=(n.base)]{\node[draw, circle, minimum size=2.9ex, inner sep=0pt, font=\scriptsize] (n) {#1};}}
\newcommand{\SCdiv}[1]{$#1$-divisible}
\newcommand{\SCtaken}[2]{\SCstep{#1} \begin{tabular}[t]{@{}l@{}} taken by \\ agent $#2$ \end{tabular}}
\newcommand{\SCrest}[2]{$#1 \setminus #2$}

\def\w{3.0}    
\def\h{1.3}     
\def\vg{1.9}    
\def\hg{0.5}    
\def\sep{0.12}  
\def\below{0.16} 

\newcommand{\SCbox}[5]{\draw[fill=#1] (#3*\w+.5*\hg,-#2*\h-#2*\vg) rectangle (#4*\w-.5*\hg,-#2*\h-#2*\vg-\h)
  node[pos=.5, align=center] {#5};}

\newcommand{\SCcut}[4]{\draw[->, >=stealth] (#2*\w,-#1*\h-#1*\vg-\h) -- (#2*\w,-#1*\h-#1*\vg-2*\h-\vg);
  \node[anchor=east, inner sep=0pt] at (#2*\w-\sep,-#1*\h-#1*\vg-\h-.5*\vg) {#3};
  \node[anchor=west, inner sep=0pt] at (#2*\w+\sep,-#1*\h-#1*\vg-\h-.5*\vg) {#4};}

\newcommand{\SCbelow}[3]{\node[anchor=north west, inner sep=0pt] at (#2*\w+.5*\hg,-#1*\h-#1*\vg-\h-\below) {#3};}

\SCbox{gray!0}{0}{0}{6}{$T$ \\ \SCdiv{161}}

\SCbox{gray!20}{1}{0}{1}{$A$ \\ \SCdiv{53}}
\SCbox{gray!0}{1}{1}{6}{$Z$ \\ \SCdiv{53}}

\SCbox{gray!20}{2}{1}{2}{$B$ \\ \SCdiv{17}}
\SCbox{gray!0}{2}{2}{6}{$C$ \\ \SCdiv{17}}

\SCbox{gray!0}{3}{2}{5}{$T'$ \\ \SCdiv{5}}
\SCbox{gray!20}{3}{5}{6}{$C^{\downarrow}$ \\ \SCdiv{5}}

\SCbox{gray!20}{4}{2}{3}{$A'$ \\ \SCdiv{1}}
\SCbox{gray!0}{4}{3}{5}{$Z'$ \\ \SCdiv{1}}

\SCbox{gray!20}{5}{3}{4}{$B'$ \\ \SCdiv{0}}
\SCbox{gray!20}{5}{4}{5}{$C'$ \\ \SCdiv{0}}

\SCcut{0}{1}{\SCstep{1} $\mathcal{T}_1(\frac{1}{3}|T)$}{\SCrest{T}{A}}
\SCcut{1}{2}{\SCstep{2} $\mathcal{T}_1(\frac{1}{3}|Z)$}{\SCrest{Z}{B}}
\SCcut{2}{5}{\SCstep{3} $\mathcal{T}_2(u_2(C)-u_2(A)|C)$}{\SCrest{C}{T'}}
\SCcut{3}{3}{\SCstep{7} $\mathcal{T}_3(\frac{1}{3} u_3(T')|T')$}{\SCrest{T'}{A'}}
\SCcut{4}{4}{\SCstep{8} $\mathcal{G}_3(\frac{1}{3} u_3(T')|Z')$}{\SCrest{Z'}{B'}}

\SCbelow{1}{0}{\SCtaken{4}{3}}
\SCbelow{2}{1}{\SCtaken{6}{1}}
\SCbelow{3}{5}{\SCtaken{5}{2}}
\SCbelow{4}{2}{\SCtaken{10}{1}}
\SCbelow{5}{3}{\SCtaken{9}{2}}
\SCbelow{5}{4}{\SCtaken{11}{3}}

\end{tikzpicture}
\caption{The imported Selfridge-Conway procedure, which constructs an envy-free allocation of the time slots into schedules for three sufficiently patient agents. As argued in the text, all details in this figure apply for any economy where (i)~each agent $i$ has a geometric utility function with discount factor $\delta_i$, (ii)~$\delta_1 \geq \frac{161}{162}$, (iii)~$\delta_2$ is sufficiently large given $\delta_1$, and (iv)~$\delta_3$ is sufficiently large given $(\delta_1, \delta_2)$. By \protect\hyperlink{Corollary3}{Corollary~3}, these conditions guarantee that the resulting allocation is envy-free.}
\end{figure*}

\vspace{\baselineskip} \noindent \textsc{Example:} {\it Imported Selfridge-Conway.} Fix a {\it sufficiently small} $\varepsilon > 0$, and {\it given $\varepsilon$} fix any three-agent economy such that (i)~each agent $i$ has a geometric utility function with discount factor $\delta_i$, (ii)~$\delta_1 \geq \frac{161}{162}$, (iii)~$\delta_2$ is {\it sufficiently large given $\delta_1$}, and (iv)~$\delta_3$ is {\it sufficiently large given $(\delta_1, \delta_2)$}. For simplicity, all italicized terms are defined implicitly by the needs of the following argument. By \hyperlink{Corollary3}{Corollary~3}, imported Selfridge-Conway delivers an envy-free allocation; the claim is that under these requirements, with agents cutting in index order, we moreover have all details in \hyperlink{Figure4}{Figure~4}. Throughout, we use without mention that a sufficiently patient agent $i$ deems each $\ell$-cycle in $S$ to be worth an amount in $[\frac{1}{\ell} \cdot u_i(S) - \varepsilon, \frac{1}{\ell} \cdot u_i(S) + \varepsilon]$, and we adjust error bounds to be precisely $\varepsilon$ using the sufficient patience requirements without mention.

First, we claim that $2$ ranks $C$ above $A$ above $B$, so that $2$ trims $C$ to construct $C^\downarrow$ that he deems as good as $A$, and we claim moreover that $2$ trims with target proportion less than $\frac{1}{2}$. Indeed, for both \textsc{Tripartition Algorithm} cuts made by agent~1, the target {\it proportion} belongs to $[\frac{1}{3}, \frac{1}{2}]$ and the value of the take pile is positive, from which it follows that some time slot in the sort pile must be skipped. Moreover, if a time slot $t$ in the sort pile is skipped and the next time slot $t'$ in the sort pile is taken, then because $\delta_1 \geq \frac{161}{162}$, (i)~$t'$ is at most nine slots later than $t$ (and nine slots later is only possible during the second cut), so (ii)~$t'$ is worth at least $90\%$ of $t$, so (iii)~after $t'$, the next time slot taken from the sort pile must be at least $100$ slots later. It follows that for each $i \in \{2, 3\}$, since $i$ is sufficiently patient given $\delta_1$, thus $i$ deems the amount that $1$ takes from either sort pile to be worth at most $\frac{1}{100} + \varepsilon$, so by sufficient patience,
\begin{itemize}
\item $u_i(A) \in [\frac{1}{3} - \varepsilon, \frac{1}{3} + \frac{1}{100} + \varepsilon]$ and $u_i(Z) \in [\frac{2}{3} - \frac{1}{100} - \varepsilon, \frac{2}{3} + \varepsilon]$,

\item $u_i(B) \in [\frac{1}{3} \cdot (\frac{2}{3} - \frac{1}{100}) - \varepsilon, \frac{1}{3} \cdot \frac{2}{3} + \frac{1}{100} + \varepsilon ] = [\frac{2}{9}- \frac{1}{300} - \varepsilon, \frac{2}{9} + \frac{1}{100} + \varepsilon]$, and

\item $u_i(C) \in [\frac{2}{3} \cdot (\frac{2}{3} - \frac{1}{100}) - \frac{1}{100} - \varepsilon, \frac{2}{3} \cdot \frac{2}{3} + \varepsilon] = [\frac{4}{9} - \frac{1}{60} - \varepsilon, \frac{4}{9} + \varepsilon]$.
\end{itemize}
Since $\varepsilon$ is sufficiently small, we have the desired ranking. Finally, the target {\it value} of the trim is $u_2(C) - u_2(A) \leq (\frac{4}{9} + \varepsilon) - (\frac{1}{3} - \varepsilon) = \frac{1}{9} + 2\varepsilon$, half the value of $C$ is $\frac{1}{2} \cdot u_2(C) \geq \frac{1}{2} \cdot (\frac{4}{9} - \frac{1}{60} - \varepsilon)$, and since $\varepsilon$ is sufficiently small, the former is less than the latter. This establishes the claim, and as a result the first case of the \textsc{Tripartition Algorithm} produces $T'$ just like it produced $A$ and $B$.

Second, we claim that $3$ selects $A$ from $\{A, B, C^\downarrow\}$. Indeed, since $3$ is sufficiently patient given $(\delta_1, \delta_2)$, and since $\varepsilon$ is sufficiently small, thus $u_3(T') \geq \frac{1}{3} \cdot u_3(C) - \varepsilon$, so $u_3(C^\downarrow) = u_3(C) - u_3(T') \leq u_3(C) - (\frac{1}{3} \cdot u_3(C) - \varepsilon) \leq \frac{8}{27} + \varepsilon < \frac{1}{3}$, and the claim follows from this and the previous paragraph.

To conclude, we claim that each $i \in \{1, 2\}$ ranks $B'$ above $A'$ above $C'$, from which all remaining details in the figure follow. Indeed, let $i \in \{1, 2\}$. For each $\delta'_3 \in (0, 1)$, let $u'_3$ denote the geometric utility function with discount factor $\delta'_3$, and define $A'(\delta'_3)$, $Z'(\delta'_3)$, $B'(\delta'_3)$, and $C'(\delta'_3)$ to be the respective versions of $A'$, $Z'$, $B'$, and $C'$ that agent~$3$ constructs from $T'$ as in \hyperlink{Figure4}{Figure~4} in the economy where we change $u_3$ to $u'_3$. Observe that if $\delta'_3$ is sufficiently large, then the arguments for the two previous paragraphs go through, so agent~$3$ indeed cuts $T'$ as in \hyperlink{Figure4}{Figure~4}. Moreover, as $\delta'_3 \to 1$, it can be shown that (i)~the first time slot taken from the sort pile when constructing $A'(\delta'_3)$ is pushed into the distant future,\footnote{This particular argument is subtle. Let $t_1$ and $t_2$ denote the first two time slots of $T'$. The cut's target proportion is $\frac{1}{3}$, and the take pile is worth more than $\frac{1}{3} u'_3(T' \backslash \{t_1, t_2\})$, which is more than $\frac{1}{3} u'_3(T') - \frac{2}{3} u'_3(\{t_1\})$, so the value required from the sort pile is less than $\frac{2}{3}$ of the value of the first time slot in $T'$, from which the claim follows.} and (ii)~the first time slot skipped when constructing $B'(\delta'_3)$ is pushed into the distant future, so (i)~$u_i(A'(\delta'_3)) \to u_i(T'_{[3|3)}) \in (0, \frac{1}{3} u_i(T')]$, (ii)~$u_i(B'(\delta'_3)) \to u_i(T'_{[1|3)} \cup T'_{[2|3)}) > \frac{2}{3} u_i(T')$, and (iii)~$u_i(C'(\delta'_3)) \to u_i(\emptyset) = 0$. Altogether, then, since $\delta_3$ is sufficiently large given $(\delta_1, \delta_2)$, thus $u_i(B') > u_i(A') > u_i(C')$, as desired.

\vspace{\baselineskip} Finally, we turn to the general case.

\hypertarget{Corollary4}{}
\vspace{\baselineskip} \noindent \textsc{Corollary~4:} Fix an economy. If for each $i \in N$ we have $u_i \in \mathcal{U}_{2 \cdot 3^{[2 \cdot 32^n \cdot n^7 \cdot (n!)^2-1]}-1} \cap \mathcal{U}_{\text{M}}$, then there is an envy-free allocation.

\vspace{\baselineskip} Indeed, according to our formal definitions, a superficial modification of the \textsc{Psomas} protocol is an envy-free cellular cake protocol whose cost given $n$ is at most $2 \cdot 32^n \cdot n^7 \cdot (n!)^2$ \citep{Psomas2026}. This is primarily driven by two critical features of the \textsc{Psomas} protocol: (i)~throughout the protocol, a \textsc{Cut} query or \textsc{Evaluate} query can only be selected if its input serving is a union of cells of the {\it current partition} determined by the previous cuts, and (ii)~throughout the protocol, a \textsc{Cut} query can only be selected if the cutter has appraised every cell of the current partition. Indeed, while the \textsc{Psomas} protocol allows a query to take as input a union of cells, the two critical features guarantee there is always an equivalent query that takes as input one cell and that can be determined using the current information. Other superficial differences include (i)~allowing for all measures with densities, which makes no difference because our technique only requires a protocol to work on a smaller domain, (ii)~allowing for arbitrary selections in the middle of the protocol (say, selecting an arbitrary agent), which intuitively makes no difference because the protocol works regardless of how these selections are made and thus works if all such selections are made in advance according to a deterministic policy, and (iii)~working with intervals of the form $[a, b)$ instead of $(a, b]$, which intuitively does not matter because cake measures assign no value to singletons. Altogether, the cost of our superficial modification is at most the worst-case number of cuts of the original, which is at most the worst-case number of queries of the original, which is given above. We remark that simple arguments could strengthen \hyperlink{Corollary4}{Corollary~4} by working with the worst-case cuts instead of the worst-case queries of the \textsc{Psomas} protocol, but we leave the current statement to facilitate verification across papers and because the improvement does not impact our main message.

To the best of our knowledge, there are currently four bounded envy-free protocols in the literature for arbitrary $n$: the \textsc{Aziz-Mackenzie} protocol \citep{Aziz-Mackenzie2016}, the \textsc{Sokolov} protocol \citep{Sokolov2023}, the \textsc{Ye-Bai} protocol \citep{Ye-Bai2026}, and the \textsc{Psomas} protocol \citep{Psomas2026}. We state a formal corollary only for the last of these, because of the four it is the only one that (i)~has a superficial modification for which it is easy to verify our formal requirements, and (ii)~provides an explicit upper bound on the worst-case number of queries. Naturally, any future protocol with these features and a tighter upper bound will immediately yield an improvement of \hyperlink{Corollary4}{Corollary~4} that requires less patience. That said, we remark that the asymptotic bound of the \textsc{Ye-Bai} protocol suggests that for sufficiently large populations, the known patience level may be such that importing a cellular modification of the \textsc{Psomas} protocol does not work but importing a cellular modification of the \textsc{Ye-Bai} protocol does.

As discussed in the introduction, while \hyperlink{Corollary4}{Corollary~4} can be interpreted as requiring a finite but astronomical level of patience to ensure there is an envy-free allocation, it can alternatively be interpreted as requiring a sufficiently fine discretization of continuous time into indivisible time slots. We remark that after completing this paper, we used frontier AI to generate an argument that the monotonicity hypothesis can be dropped, and the argument appears sound to us; see \hyperlink{Section4}{Section~4} for details.

\hypertarget{Section4}{}
\section{Discussion}

We have shown that (i)~if agents are sufficiently patient, then there are proportional allocations, and (ii)~if agents are sufficiently patient and moreover have weak preference for earlier time slots, then there are envy-free allocations.

In order to establish the former result, we showed that whenever agents have $(2n-3)$-Kakeya preferences, we can map the given economy to an associated economy with monotonic preferences and run \textsc{Iterative Cycle Apportionment}, then use this allocation to determine the order in which agents select time slots in the original economy. We remark that this argument works whenever the agents {\it in the associated monotonic economy} agree that each $(n-1)$-cycle of $T$ is $1$-divisible, and that more elaborate arguments could improve our bound on the sufficient level of patience for this argument to go through, but we have chosen to omit such arguments as they would have little impact on our main message.\footnote{In fact, since our initial draft, this very question has been investigated by Christoph Schwerdtfeger, who reports to us via personal communication that sharpening the dense-subset divisibility estimate can lower the required patience to a level given implicitly by a particular equation; for $n=4$, Christoph reports that this improves our requirement of $5$ to approximately $4.49$.} Indeed, this argument cannot establish the conjecture of \cite{Mackenzie-Komornik2023} that there are proportional allocations whenever agents have $(n-1)$-Kakeya preferences, and there is no particular reason to believe that this argument establishes the tightest possible bound. That said, it does suffice to establish that a bound on the level of patience that is linear in the number of agents guarantees that there are proportional allocations, and moreover this bound is tight when $n=2$.

In order to establish the latter result, we introduced the \textsc{Tripartition Algorithm} that allows an agent to divide a set that everybody deems divisible into two sets such that (i)~he deems one of them to be worth a target value, and (ii)~everybody agrees that both sets retain some divisibility. Because a superficial modification of the \textsc{Psomas} protocol \citep{Psomas2026} is an envy-free cellular cake protocol with finite cost, we can import it to our setting using the \textsc{Tripartition Algorithm} and \textsc{Greedy Algorithm} to make cuts, provided the agents are sufficiently patient and have weak preference for earlier time slots. In fact, the same logic applies to any other cake protocols with these features. Unfortunately, however, this argument has its limitations: in order to use the \textsc{Tripartition Algorithm} to import {\it any} protocol that ultimately constructs an allocation by making at least $n-1$ cuts, we require a level of patience that is at least exponential in the number of agents.\footnote{We remark that there is a lower bound on the total number of queries required to construct an envy-free allocation \citep{Procaccia2009}, but because this does not yield a lower bound on the number of cuts, we cannot use it to draw stronger conclusions about the limitations of our argument for no-envy. That said, bounds on the number of required cuts have been previously investigated in the context of proportionality (\citealp{Even-Paz1984}; \citealp{Edmonds-Pruhs2006}), and analogous future work for no-envy could indeed provide further insight about this.} That said, as discussed in the introduction, an alternative interpretation is that instead of requiring the agents to be sufficiently patient, we require continuous time to have a sufficiently fine partition into time slots in order to guarantee there is an envy-free allocation.

We remark that after completing this paper, we used frontier AI to generate an argument that monotonicity is not required for the latter result: so long as agents are sufficiently patient, there are envy-free allocations. Though the AI-generated argument looks sound to us, it is substantially more complicated than the one that we developed. At a high level, the idea is to generalize our \textsc{Tripartition Algorithm} so that instead of using cycles to construct the sort-skip-take schedules, we use the entire utility function profile and a technique due to \cite{Barany-Doerr2006} to construct the sort-skip-take schedules. More precisely, the B\'{a}r\'{a}ny-Doerr technique is used to take an input schedule that all agree is `very divisible' and a target proportion, then output a sub-schedule of the input schedule with the property that for each input tail, there is consensus that the proportion of value taken by the output is approximately the target proportion, with this approximation relative to the value of the input tail. This in turn is used to construct the sort-skip-take schedules to generalize our \textsc{Tripartition Algorithm} in a manner that does not require monotonicity, at which point we can import cake protocols as in this paper. In this way, the later AI-generated improvement builds on the simpler argument presented here.

We conclude by highlighting an open question: are there envy-free allocations that are moreover efficient whenever agents are sufficiently patient?

\appendix
\hypertarget{Appendix1}{}
\setcounter{secnumdepth}{0}
\section{Appendix 1}

In this appendix, we prove \hyperlink{Lemma1}{Lemma~1}.

\vspace{\baselineskip} \noindent \textsc{Lemma~1 (Restated):} Fix an economy. For each $i \in N$ such that $u_i \in \mathcal{U}_{\text{M}}$, each $k \in [0, \infty)$, each $S \in \mathcal{S}$ that is $k$-divisible for~$i$, each $\ell \in \mathbb{N}$ such that $\ell \leq k+1$, and each $S^* \subseteq S$ that is $\ell$-dense in $S$, we have that $S^*$ is $[(\frac{1}{\ell}) \cdot (k - (\ell-1))]$-divisible for $i$.

\vspace{\baselineskip} \noindent \textsc{Proof:} Assume the hypotheses. If $k=0$, then we are done; thus let us assume $k>0$. Since $S^*$ is $\ell$-dense in $S$, thus $S$ is infinite and $S^*$ contains an infinite cycle. Let $r^* \in \mathbb{N}$ be an arbitrary rank for use in $S^*$, and let $r \in \mathbb{N}$ be the associated rank for use in $S$ defined by $S_r = S^*_{r^*}$. Since~$S^*$ is $\ell$-dense in $S$, thus $S^*_{[r^*+1)}$ is as well. To proceed, we establish four inequalities.
\begin{itemize}
\item First, since (i)~by the $\ell$-density of $S^*_{[r^*+1)}$ in $S$, the earliness-preserving bijection from $S^*_{[r^*+1)}$ to $S_{[r+\ell|\ell)}$ maps each time slot to one that is at least as late, and (ii)~$u_i \in \mathcal{U}_{\text{M}}$, thus $u_i(S^*_{[r^*+1)}) \geq u_i(S_{[r+\ell|\ell)})$.

\item Second, since (i)~$S_{[r+\ell)}$ is partitioned into $\ell$ cycles of length $\ell$, of which $S_{[r+\ell|\ell)}$ has the earliest starting time slot, (ii)~for each pair $r_0, r_1 \in \mathbb{N}$ such that $r_0 < r_1$, the earliness-preserving bijection from $S_{[r_0|\ell)}$ to $S_{[r_1|\ell)}$ maps each time slot to a later one, and (iii)~$u_i \in \mathcal{U}_{\text{M}}$, thus $u_i(S_{[r+\ell|\ell)}) \geq (\frac{1}{\ell}) \cdot u_i(S_{[r+\ell)})$.

\item Third, since (i)~$S_{[r+\ell)} = S_{[r+1)} \setminus \{\tau_{r+r'}(S)\}_{r' \in \llbracket 1, \ell-1 \rrbracket}$, and (ii)~$u_i \in \mathcal{U}_{\text{M}}$, thus $u_i(S_{[r+\ell)}) \geq u_i(S_{[r+1)}) - (\ell - 1) \cdot u_i(S_r) = u_i(S_{[r+1)}) - (\ell - 1) \cdot u_i(S^*_{r^*})$.

\item Finally, since $S$ is $k$-divisible for $i$, thus $u_i(S_{[r+1)}) \geq k \cdot u_i(S_r) = k \cdot u_i(S^*_{r^*})$.
\end{itemize}
By the above inequalities, $u_i(S^*_{[r^*+1)}) \geq [(\frac{1}{\ell}) \cdot (k - (\ell-1))] \cdot u_i(S^*_{r^*})$. Since $r^* \in \mathbb{N}$ was arbitrary, thus $S^*$ is $[(\frac{1}{\ell}) \cdot (k - (\ell - 1))]$-divisible for $i$, as desired.~$\blacksquare$

\hypertarget{Appendix2}{}
\setcounter{secnumdepth}{0}
\section{Appendix 2}

In this appendix, we prove \hyperlink{Lemma2}{Lemma~2}, \hyperlink{Lemma3}{Lemma~3}, and \hyperlink{Theorem1}{Theorem~1}. We begin with \hyperlink{Lemma2}{Lemma~2}.

\vspace{\baselineskip} \noindent \textsc{Lemma~2 (Restated):} 
For each $k \in [0,\infty)$ and each $u_0 \in \mathcal{U}_k$, $u_0^{\text{M}} \in\mathcal{U}_k \cap \mathcal{U}_{\text{M}}$.

\vspace{\baselineskip} \noindent \textsc{Proof:} By construction, we have $u_0^{\text{M}}\in\mathcal{U}_{\text{M}}$; thus we need only show $u_0^{\text{M}}\in\mathcal{U}_k$. If $k=0$, then we are done; thus let us assume $k > 0$. Let $T^+ \subseteq T$ denote the set of time slots to which $u_0$ assigns positive value.

First, we claim that $T^+$ is infinite. Indeed, (i)~$u_0$ cannot assign positive value to zero time slots, as this would contradict $\sum_{t \in T} u_0(\{t\}) = 1$; and (ii)~$u_0$ cannot assign positive value to a positive finite number of time slots, as otherwise there would be a final time slot with positive value, contradicting that $u_0 \in \mathcal{U}_k$ with $k>0$; thus we must have the desired claim.

Second, we claim there is a bijection $\beta: T^+ \to T$ such that for each $t \in T^+$, $u_0(\{t\}) = u_0^{\text{M}}(\{\beta(t)\})$. Indeed, we can construct such a $\beta$ explicitly: start with all of $T^+$, and then iteratively (i)~let $v$ denote the highest value assigned by $u_0$ to a remaining time slot, which is well-defined because an increasing sequence of values would contradict $\sum_{t \in T} u_0(\{t\}) = 1$, (ii)~let $t$ denote the earliest remaining time slot to which $u_0$ assigns $v$, (iii)~define $\beta(t)$ to be the earliest time slot in $T$ that has not yet been assigned by $\beta$, and (iv)~remove $t$ from the set of remaining time slots. Since $\sum_{t \in T} u_0(\{t\}) = 1$, thus for each $t \in T^+$ we have that $u_0$ can only assign a value that is at least $u_0(\{t\})$ to a finite collection of time slots, so each time slot in $T^+$ is eventually removed, and using the construction of $u_0^{\text{M}}$ it is straightforward to verify that $\beta$ satisfies the desired properties. Observe that by this claim, $u_0^{\text{M}}$ assigns positive value to {\it all} time slots. For the rest of the proof, let $\beta$ denote the constructed bijection.

To conclude, let $t \in T$ and define $t^* \equiv \max\{t'\in T\mid u_0(\{t'\}) \geq u_0^{\text{M}}(\{t\})\}$; this is well-defined because (i)~as observed above, $u_0^{\text{M}}$ assigns positive value to all time slots, including~$t$, and (ii)~$\sum_{t' \in T} u_0(\{t'\}) = 1$. Let us now consider $T^+_{\rightarrow} \equiv T^+ \cap \{t^*+1, t^*+2, \dots\}$ for $u_0$ while we consider $\beta(T^+_{\rightarrow}) \equiv \{\beta(t') | t' \in T^+_{\rightarrow}\}$ for $u_0^{\text{M}}$. Indeed, observe that (i)~by construction of $t^*$, $u_0$ assigns each time slot in $T^+_{\rightarrow}$ a value below $u_0^{\text{M}}(\{t\})$, (ii)~by the second claim, $u_0(T^+_{\rightarrow}) = u_0^{\text{M}}(\beta(T^+_{\rightarrow}))$, and (iii)~because $u_0^{\text{M}}$ is monotonic, and because by definition of $\beta$ we have that $u_0^{\text{M}}$ assigns each time slot in $\beta(T^+_{\rightarrow})$ a value below $u_0^{\text{M}}(\{t\})$, thus $\beta(T^+_{\rightarrow}) \subseteq \{t+1, t+2, \dots\}$; altogether, then, $u_0^{\text{M}}(\{t+1,t+2,\dots\}) \geq u_0^{\text{M}}(\beta(T^+_{\rightarrow})) = u_0(T^+_{\rightarrow}) = u_0(\{t^*+1,t^*+2,\dots\})$. By (i)~the preceding inequality; (ii)~since $u_0 \in \mathcal{U}_k$, we have $u_0(\{t^*+1,t^*+2,\dots\}) \geq k \cdot u_0(\{t^*\})$; and (iii)~by definition of $t^*$, we have $u_0(\{t^*\}) \geq u_0^{\text{M}}(\{t\})$; thus altogether we have that $u_0^{\text{M}}(\{t+1,t+2,\dots\}) \geq k \cdot u_0^{\text{M}}(\{t\})$. Since $t \in T$ was arbitrary, we are done.~$\blacksquare$

\vspace{\baselineskip} Next, we prove \hyperlink{Lemma3}{Lemma~3}.

\vspace{\baselineskip} \noindent \textsc{Lemma~3 (Restated):} For each economy $(n, u)$, if $(n, u^{\text{M}})$ has a proportional allocation, then $(n, u)$ does as well.

\vspace{\baselineskip} \noindent \textsc{Proof:} 
Let $(n,u)$ be an economy and assume that $\pi^{\text{M}} \in \Pi$ is proportional for $(n,u^{\text{M}})$. We construct $\pi \in \Pi$ as follows. First, for each $t \in T$, let $i_t$ denote the agent who receives~$t$ according to $\pi^{\text{M}}$. Second, we proceed through the time slots in sequence, and at each time slot $t$, agent $i_t$ consumes the remaining time slot $f(t)$ that (i)~is at least as valuable as all other remaining time slots according to $u_{i_t}$, and (ii)~is earliest among such time slots. Finally, agent $1$ consumes any remaining time slots.

We claim that $\pi$ is proportional for $(n, u)$. Indeed, let $i \in N$, let $t \in \pi^{\text{M}}_i$, and let $T^* \subseteq T$ be a set of $t-1$ time slots to which $u_i$ assigns the highest value. Then
\begin{align*}
u_i(\{f(t)\}) &= \max_{t'' \in T \setminus \{f(t') \mid t' \in \llbracket 1, t-1 \rrbracket\}} u_i(\{t''\})
\\ &\geq \max_{t' \in T \setminus T^*} u_i(\{t'\})
\\ &= u^{\text{M}}_i(\{t\}).
\end{align*}
Since $t \in \pi^{\text{M}}_i$ was arbitrary, thus $u_i(\pi_i) \geq u^{\text{M}}_i(\pi^{\text{M}}_i) \geq \frac{1}{n}$. Since $i \in N$ was arbitrary, we are done.~$\blacksquare$

\vspace{\baselineskip} To conclude this appendix, we prove \hyperlink{Theorem1}{Theorem~1}.

\vspace{\baselineskip} \noindent \textsc{Theorem~1 (Restated):} Fix an economy with at least two agents. If for each $i \in N$ we have $u_i \in \mathcal{U}_{2n-3}$, then there is a proportional allocation.

\vspace{\baselineskip} \noindent \textsc{Proof:} Assume the hypotheses. By \hyperlink{Lemma2}{Lemma~2}, for each $i \in N$ we have $u^{\text{M}}_i \in \mathcal{U}_{2n-3} \cap \mathcal{U}_{\text{M}}$, and thus $u^{\text{M}}_i$ assigns a positive utility to each time slot.

We claim that $(n, u^{\text{M}})$ has a proportional allocation. Indeed, we construct the desired allocation using \textsc{Iterative Cycle Apportionment}, which consists of $n$ stages indexed by $\llbracket 1, n \rrbracket$. In each stage $\mathsf{s} \in \llbracket 1, n-1 \rrbracket$, the $n-(\mathsf{s}-1)$ remaining agents consider the remaining time slots in the top $\mathsf{s}$ $(n-1)$-cycles in order to fill a basket and assign its contents $B^\mathsf{s}$ to an agent $i^\mathsf{s}$. In the final stage, the remaining time slots are assigned to the remaining agent.

For each stage $\mathsf{s} \in \llbracket 1, n-1 \rrbracket$, let (i)~$N^\mathsf{s} \equiv N \setminus \{ i^{\mathsf{s}'} \in N \mid \mathsf{s}' < \mathsf{s} \}$ denote the set of remaining agents at the start of stage~$\mathsf{s}$, (ii)~$T^\mathsf{s} \equiv T \setminus [\cup_{\mathsf{s}' < \mathsf{s}} B^{\mathsf{s}'}]$ denote the remaining time slots at the beginning of stage $\mathsf{s}$, and (iii)~$C^\mathsf{s} \equiv T^\mathsf{s} \cap [\cup_{\mathsf{s}' \leq \mathsf{s}} T_{[\mathsf{s}'|n-1)}]$ denote the time slots under consideration during stage~$\mathsf{s}$. Before providing further details about the construction of $(B^\mathsf{s})_{\mathsf{s} \in \llbracket 1, n-1 \rrbracket}$, we make two observations.

First, we observe that for each stage $\mathsf{s} \in \llbracket 1, n-1 \rrbracket$ and each $i \in N$, $C^\mathsf{s}$ is $1$-divisible for~$i$. Indeed, since $u^{\text{M}}_i \in \mathcal{U}_{2n-3}$, thus $T$ is $(2n-3)$-divisible for $i$. Moreover, since $T_{[\mathsf{s}|n-1)} \subseteq C^\mathsf{s}$ and $\mathsf{s} \in \llbracket 1, n-1 \rrbracket$, thus $C^\mathsf{s}$ is $(n-1)$-dense in $T$. Finally, $u^{\text{M}}_i \in \mathcal{U}_{\text{M}}$. Altogether, then, by \hyperlink{Lemma1}{Lemma~1} we have that $C^\mathsf{s}$ is $[(\frac{1}{n-1})\cdot((2n-3)-(n-2))]$-divisible for $i$, as desired.

Second, we observe that for each stage $\mathsf{s} \in \llbracket 1, n-1 \rrbracket$ and each $i \in N$, we have $u^{\text{M}}_i(C^\mathsf{s}) \geq (\frac{1}{n-(\mathsf{s}-1)}) \cdot u^{\text{M}}_i(T^\mathsf{s})$. Indeed, define $C^* \equiv \cup_{\mathsf{s}' \in \llbracket \mathsf{s}, n-1 \rrbracket} T_{[\mathsf{s}'|n-1)}$. Since $u_i^{\text{M}} \in \mathcal{U}_{\text{M}}$, thus
\begin{align*}
u^{\text{M}}_i(C^\mathsf{s}) &= u^{\text{M}}_i(T^\mathsf{s} \setminus C^*) + u^{\text{M}}_i(T_{[\mathsf{s}|n-1)})
\\ &\geq u^{\text{M}}_i(T^\mathsf{s} \setminus C^*) + \Bigl(\frac{1}{n-(\mathsf{s}-1)}\Bigr) \cdot u^{\text{M}}_i(C^*)\\
&=  u^{\text{M}}_i(T^\mathsf{s} \setminus C^*) + \Bigl(\frac{1}{n-(\mathsf{s}-1)}\Bigr) \cdot u^{\text{M}}_i(T^\mathsf{s} \cap C^*)\\
&= \Bigl[1 \cdot \Bigl(\frac{u^{\text{M}}_i(T^\mathsf{s} \setminus C^*)}{u^{\text{M}}_i(T^\mathsf{s})}\Bigr) + \Bigl(\frac{1}{n-(\mathsf{s}-1)}\Bigr) \cdot \Bigl(\frac{u^{\text{M}}_i(T^\mathsf{s} \cap C^*)}{u^{\text{M}}_i(T^\mathsf{s})}\Bigr)\Bigr] \cdot u^{\text{M}}_i(T^\mathsf{s}) \\
&\geq \Bigl(\frac{1}{n-(\mathsf{s}-1)}\Bigr) \cdot u^{\text{M}}_i(T^\mathsf{s}),
\end{align*}
as desired. To ease verification, we highlight that the five lines above use (i)~the definition of $C^\mathsf{s}$, (ii)~the fact that $u^{\text{M}}_i \in \mathcal{U}_{\text{M}}$, (iii)~the fact that $C^* \subseteq T^\mathsf{s}$, (iv)~simple re-writing, and (v)~the fact that $u_i^{\text{M}}(T^\mathsf{s} \setminus C^*) + u^{\text{M}}_i(T^\mathsf{s} \cap C^*) = u^{\text{M}}_i(T^\mathsf{s})$, respectively.

We now describe how baskets are filled and assigned. In each stage $\mathsf{s} \in \llbracket 1, n-1 \rrbracket$, the $n-(\mathsf{s}-1)$ remaining agents consider the members of $C^\mathsf{s}$ in sequence. At each time slot~$t$, each agent places a flag in the basket if and only if he measures the value of the basket with $t$ to exceed $\frac{1}{n}$. If there are no flags, then $t$ is added to the basket and the remaining agents move to the next time slot. If there is one flag, then let $i^\mathsf{s}$ be the agent who placed the flag, let $B^\mathsf{s}$ denote the contents of the basket with $t$, and assign $B^\mathsf{s}$ to~$i^\mathsf{s}$; in this case the stage ends and we move to the next stage. If there are multiple flags, then~$t$ is skipped and the agents move to the next time slot. If all time slots are considered and there is no time slot with one flag, then let $B^\mathsf{s}$ denote the basket's limit schedule (that is, the union of its schedules across all time periods), let $i^\mathsf{s}$ be any remaining agent whose utility for~$B^\mathsf{s}$ is highest, and assign $B^\mathsf{s}$ to $i^\mathsf{s}$.

We claim that for each $\mathsf{s} \in \llbracket 1, n-1 \rrbracket$, (i)~$u^{\text{M}}_{i^\mathsf{s}}(B^\mathsf{s}) \geq \frac{1}{n}$, and (ii)~$i \in N^\mathsf{s} \setminus \{i^\mathsf{s}\}$ implies $u^{\text{M}}_i(B^\mathsf{s}) \leq \frac{1}{n}$. We proceed by induction, handling the base step and inductive step together. Indeed, assume that for each $\mathsf{s}' < \mathsf{s}$ we have both statements. By construction, we have the second statement for $\mathsf{s}$. If there is a time slot with one flag, then we are done; thus let us assume there is no such time slot. By the inductive hypothesis, for each $i \in N^\mathsf{s}$ we have $u^{\text{M}}_i(T^\mathsf{s}) \geq \frac{n-(\mathsf{s}-1)}{n}$, so by the second observation we have $u^{\text{M}}_i(C^\mathsf{s}) \geq \frac{1}{n}$. If each time slot has no flag, then we are done; thus let us assume there is a time slot with multiple flags. By the first observation, for each $i \in N^\mathsf{s}$ we have that $C^\mathsf{s}$ is $1$-divisible for~$i$; this implies that there cannot be a maximum time slot with multiple flags by a straightforward argument, which we include in the following footnote for completeness.\footnote{Assume by way of contradiction that in stage $\mathsf{s}$, (i)~there is no time slot with one flag, and (ii)~$t^*$ is the maximum time slot with multiple flags. Let $i \in N^\mathsf{s}$ be an agent who places a flag at $t^*$ in stage $\mathsf{s}$. Since $C^\mathsf{s}$ is $1$-divisible for $i$, thus $i$ deems the later time slots in $C^\mathsf{s}$ to be worth at least $t^*$ alone, so $i$ deems the basket at $t^*$ plus all later time slots to be worth too much, so by countable additivity of $u^{\text{M}}_i$ there must be a later time slot at which $i$ places a flag, contradicting that there are no time slots with one flag and no later time slots with multiple flags. We remark that an analogous argument appears in the proof of Theorem~6 in \cite{Mackenzie-Komornik2023}.} Then there is $i \in N^\mathsf{s}$ who places an infinite collection of flags. Since $\lim_{t \to \infty} u_i^{\text{M}}(\{t\}) = 0$, thus $u^{\text{M}}_i(B^\mathsf{s}) \geq \frac{1}{n}$, so $u^{\text{M}}_{i^\mathsf{s}}(B^\mathsf{s}) \geq \frac{1}{n}$, as desired.

By the above claim, \textsc{Iterative Cycle Apportionment} constructs a proportional allocation for $(n, u^{\text{M}})$, so by \hyperlink{Lemma3}{Lemma~3} we have that $(n, u)$ has a proportional allocation, as desired.~$\blacksquare$

\hypertarget{Appendix3}{}
\setcounter{secnumdepth}{0}
\section{Appendix 3}

In this appendix, we prove \hyperlink{Proposition1}{Proposition~1} and \hyperlink{Theorem2}{Theorem~2}. We begin with \hyperlink{Proposition1}{Proposition~1}.

\vspace{\baselineskip} \noindent \textsc{Proposition~1 (Restated):} Fix an economy. For each pair $i, j \in N$ such that $u_i \in \mathcal{U}_{\text{M}}$ and $u_j \in \mathcal{U}_{\text{M}}$, each $k \in [5, \infty)$, each $S \in \mathcal{S}$ that is $k$-divisible for both $i$ and $j$, and each $v \in (0, u_i(S))$, if $S^* = \mathcal{T}_i(v|S)$, then
\begin{itemize}
\item $u_i(S^*) = v$,

\item $S^*$ is $[(\frac{1}{3}) \cdot (k - 2)]$-divisible for both $i$ and $j$, and

\item $S \setminus S^*$ is $[(\frac{1}{3}) \cdot (k - 2)]$-divisible for both $i$ and $j$.
\end{itemize}

\vspace{\baselineskip} \noindent \textsc{Proof:} Assume the hypotheses. If we establish that $v \leq (\frac{1}{2}) \cdot u_i(S)$ implies the desired conclusion, then by construction we are done; thus let us assume $v \leq (\frac{1}{2}) \cdot u_i(S)$. Define $r$, $S^{\text{sort}}$, $S^{\text{skip}}$, and $S^{\text{take}}$ as in the construction of $\mathcal{T}_i(v|S) = S^*$; thus $(\frac{1}{3}) \cdot u_i(S_{[r)}) \leq v$.

First, we claim that $u_i(S^{\text{skip}}) \leq (\frac{1}{2}) \cdot u_i(S_{[r)})$. Indeed, since $S_{[r+1|3)}=S^{\text{skip}}$ and $u_i \in \mathcal{U}_{\text{M}}$, thus $u_i(S_{[r)}) \geq u_i(S_{[r|3)}) + u_i(S^{\text{skip}})$ and $u_i(S_{[r|3)}) \geq u_i(S^{\text{skip}})$. Altogether, then, $u_i(S^{\text{skip}}) \leq (\frac{1}{2}) \cdot u_i(S_{[r)})$, as desired.

Second, we claim that $v \leq (\frac{1}{2}) \cdot u_i(S_{[r)})$. Indeed, if $r=1$, then $S_{[r)} = S$ and we are done; thus let us assume $r>1$. Since $S$ is $5$-divisible for $i$ we have $u_i(S_{[r)}) \geq 5 \cdot u_i(S_{r-1})$, and by construction of $r$ we have $(\frac{1}{3}) \cdot u_i(S_{[r-1)}) > v$, so altogether we have
\begin{align*}
\Bigl(\frac{1}{2} \Bigr) \cdot u_i(S_{[r)}) &= \Bigl(\frac{1}{2}\Bigr) \cdot \Bigl( \frac{u_i(S_{[r)})}{u_i(S_{r-1}) + u_i(S_{[r)})} \Bigr) \cdot u_i(S_{[r-1)})
\\ &\geq \Bigl(\frac{1}{2}\Bigr) \cdot \Bigl(\frac{5}{6}\Bigr) \cdot u_i(S_{[r-1)})
\\ &> \Bigl(\frac{1}{3}\Bigr) \cdot u_i(S_{[r-1)})
\\ &> v,
\end{align*}
as desired.

Third, we claim that $u_i(S^{\text{take}}) \leq (\frac{1}{3}) \cdot u_i(S_{[r)})$. Indeed, since $S_{[r+2|3)}=S^{\text{take}}$ and $u_i \in \mathcal{U}_{\text{M}}$, thus $u_i(S_{[r)}) = u_i(S_{[r|3)}) + u_i(S_{[r+1|3)}) + u_i(S^{\text{take}})$, $u_i(S_{[r|3)}) \geq u_i(S^{\text{take}})$, and $u_i(S_{[r+1|3)}) \geq u_i(S^{\text{take}})$. Altogether, then, $u_i(S^{\text{take}}) \leq (\frac{1}{3}) \cdot u_i(S_{[r)})$, as desired.

Fourth, we claim that $S^{\text{sort}}$ is $1$-divisible for $i$. Indeed, since (i)~$S$ is $5$-divisible for $i$, and (ii)~$S^{\text{sort}}$ is $3$-dense in $S$, thus by \hyperlink{Lemma1}{Lemma~1} we have that $S^{\text{sort}}$ is $[(\frac{1}{3})\cdot(5-2)]$-divisible for $i$, as desired.

To conclude, by the first two claims we have $u_i(S^{\text{sort}})+ u_i(S^{\text{take}}) = u_i(S_{[r)}) - u_i(S^{\text{skip}}) \geq (\frac{1}{2}) \cdot u_i(S_{[r)}) \geq v$, so $u_i(S^{\text{sort}}) \geq v - u_i(S^{\text{take}})$. Moreover, by construction of $r$ and the third claim we have $v - u_i(S^{\text{take}}) \geq (\frac{1}{3}) \cdot u_i(S_{[r)}) - (\frac{1}{3}) \cdot u_i(S_{[r)}) = 0$, so altogether we have $v - u_i(S^{\text{take}}) \in [0, u_i(S^{\text{sort}})]$. Finally, by the fourth claim, $S^{\text{sort}}$ is $1$-divisible for $i$. Altogether, then, by \hyperlink{TheoremMK}{Theorem~MK} we have that $u_i(\mathcal{G}_i(v - u_i(S^{\text{take}})|S^{\text{sort}})) = v - u_i(S^{\text{take}})$, so $u_i(S^*) = v$. Finally, (i)~$u_i \in \mathcal{U}_{\text{M}}$ and $u_j \in \mathcal{U}_{\text{M}}$; (ii)~$S$ is $k$-divisible for both $i$ and~$j$; and (iii)~since $S_{[r+2|3)} \subseteq S^* \subseteq S_{[r)}$ and $(S \setminus S_{[r)}) \cup S_{[r+1|3)} \subseteq S \setminus S^*$, thus $S^*$ and $S \setminus S^*$ are both $3$-dense in $S$; thus by \hyperlink{Lemma1}{Lemma~1}, we have that $S^*$ and $S \setminus S^*$ are both $[(\frac{1}{3})\cdot(k-(3-1))]$-divisible for both $i$ and~$j$, as desired.~$\blacksquare$

\vspace{\baselineskip} To conclude this appendix, we prove \hyperlink{Theorem2}{Theorem~2}.

\vspace{\baselineskip} \noindent \textsc{Theorem~2 (Restated):} Fix an economy. For each number of cuts $c \in \mathbb{N}$, if (i)~the associated cake setting has an envy-free cellular cake protocol $\mathsf{Q}$ such that $\mathsf{cost}(\mathsf{Q}) \leq c$, and (ii)~for each $i \in N$ we have $u_i \in \mathcal{U}_{\mathsf{d}(c)} \cap \mathcal{U}_{\text{M}}$, then there is an envy-free allocation.

\vspace{\baselineskip} \noindent \textsc{Proof:} Assume the hypotheses are satisfied by economy $(n, u)$, the cut quantity $c \in \mathbb{N}$, and the cake protocol $\mathsf{Q}$. Throughout this proof, we refer to an allocation in $\Pi$ as a {\it time allocation}, to distinguish from cake allocations.

\vspace{\baselineskip} \noindent \textsc{Step 1:} Introduce preliminary concepts.

\vspace{\baselineskip} First, an {\it appraised cake partition} is a pair $(\mathcal{P}^C, (\alpha^C_i)_{i \in N})$, where (i)~$\mathcal{P}^C$ is a partition of $C$ called the {\it cake partition}, and (ii)~for each $i \in N$, $\alpha^C_i: \mathcal{P}^C \to [0, 1]$ is a function that associates each cell of the cake partition with an evaluation. We write $(\mathcal{P}^C, \alpha^C)$ for brevity, and we refer to each member of $\mathcal{P}^C$ as a {\it cake cell}.

Second, an {\it appraised time partition} is a pair $(\mathcal{P}^T, (\alpha^T_i)_{i \in N})$, where (i)~$\mathcal{P}^T$ is a partition of $T$ called the {\it time partition}, and (ii)~for each $i \in N$, $\alpha^T_i: \mathcal{P}^T \to [0, 1]$ is a function that associates each cell of the time partition with an evaluation. We write $(\mathcal{P}^T, \alpha^T)$ for brevity, and we refer to each member of $\mathcal{P}^T$ as a {\it time cell}.

Third, we construct an envy-free time allocation through a procedure we call the {\it time protocol}, to distinguish it from the cake protocol. The time protocol iteratively uses $\mathsf{Q}$ to cut both $C$ and $T$, and at each stage the relevant information is summarized by a {\it construction state}, or a tuple $(I, \mathcal{P}^C, \alpha^C, \mathcal{P}^T, \alpha^T, \mathsf{m})$ defined as follows.
\begin{itemize}
\item $I$ is an information set of the cake protocol called the {\it cake infoset}. Formally, this is identified with a {\it chronicle}, or an ordered list of queries and responses; see \hyperlink{Appendix4}{Appendix~4} for details.

\item $(\mathcal{P}^C, \alpha^C)$ is an appraised cake partition.

\item $(\mathcal{P}^T, \alpha^T)$ is an appraised time partition, with $|\mathcal{P}^T| = |\mathcal{P}^C|$.

\item $\mathsf{m}: \mathcal{P}^C \to \mathcal{P}^T$ is a bijection called the {\it cell matching}.
\end{itemize}
Throughout the time protocol, the construction state will satisfy some further consistency conditions---for example, each agent's appraisal of a cake cell will be equal to his appraisal of its matched time cell---but rather than provide a thorough inventory of these conditions at this point, we use this simple definition of construction state to define our time protocol and then we study the properties of our procedure. The final output of the time protocol is an envy-free time allocation.

\vspace{\baselineskip} \noindent \textsc{Step 2:} Define the time protocol.

\vspace{\baselineskip} The initial construction state $(I, \mathcal{P}^C, \alpha^C, \mathcal{P}^T, \alpha^T, \mathsf{m})$ is such that (i)~$I$ is the initial information set of the mediator in $\mathsf{Q}$, (ii)~$\mathcal{P}^C = \{C\}$, and for each $i \in N$ we have $\alpha^C_i(C) = 1$, (iii)~$\mathcal{P}^T = \{T\}$, and for each $i \in N$ we have $\alpha^T_i(T) = 1$, and (iv)~$\mathsf{m}$ maps $C$ to $T$.

Recycling notation, let $(I, \mathcal{P}^C, \alpha^C, \mathcal{P}^T, \alpha^T, \mathsf{m})$ now denote an arbitrary construction state reached by our time protocol. Given the construction state, the time protocol may either (i)~update the construction state to $(I^+, \mathcal{P}^{C+}, \alpha^{C+}, \mathcal{P}^{T+}, \alpha^{T+}, \mathsf{m}^+)$, or (ii)~terminate with a time allocation. We describe the two cases below.

\vspace{\baselineskip} \noindent \textsc{Case 1:} $\mathsf{Q}(I)$ is a cake allocation. In this case, we terminate the time protocol with an associated time allocation. If the cake allocation simply assigns cake partition cells to agents, then the time allocation simply assigns each time partition cell to the agent that receives the matching cake partition cell.

In principle, the cake allocation need not respect the cake partition, even though intuitively the mediator should never find it useful to make cuts without asking the agents to evaluate the new pieces. This intuition is correct: we claim that if there is a cake allocation that is guaranteed to be envy-free, in the sense that it is $\mu$-envy-free for each measure profile $\mu$ that is compatible with $I$, then there is such a cake allocation that moreover assigns each agent a union of cake cells. In this case, we select such a cake allocation in a particular way, and then again terminate with the time allocation that simply assigns each time partition cell to the agent that receives the matching cake partition cell. The rest of the Case~1 argument is dedicated to proving our claim and specifying our particular selection; these details are not essential to understanding the main ideas of the proof, but are included for completeness.

Formally, let $X$ be the cake allocation selected by $\mathsf{Q}$ at $I$. Since $\mathsf{Q}$ is envy-free, thus for each cake measure profile $\mu$ compatible with $I$, $X$ is $\mu$-envy-free. For each cake cell $S \in \mathcal{P}^C$, let $i(S)$ denote the lowest-index agent $i$ such that $X_i \cap S \neq \emptyset$, and define $X'$ by assigning each cake cell $S$ to $i(S)$. Then $X'$ assigns cake cells to agents, and we claim that it is also envy-free for all compatible cake measure profiles. Indeed, for each $i \in N$ and each $S \subseteq C$ that is a union of cells of $\mathcal{P}^C$, in a slight abuse of notation to extend the domain of $\alpha^C_i$, let us write $\alpha^C_i(S) \equiv \sum_{S' \in \mathcal{P}^C | S' \subseteq S} \alpha^C_i(S')$. For each $\varepsilon \in (0, 1)$, let $\mu^\varepsilon$ be the cake measure profile such that for each $i \in N$ and each cake cell $S \in \mathcal{P}^C$, we have the following.
\begin{itemize}
\item If $S \cap X_{i(S)} = S$, then the value $\alpha^C_i(S)$ is distributed entirely to $S \cap X_{i(S)}$.

\item If $S \cap X_{i(S)} \subsetneq S$, then the value $\alpha^C_i(S)$ is distributed by assigning $(1-\varepsilon) \cdot \alpha^C_i(S)$ to $S \cap X_{i(S)}$ and distributing the remaining value evenly among the nonempty members of $\{S \cap X_j\}_{j \in N \backslash \{i(S)\}}$.

\item Within each nonempty serving in $\{S \cap X_j\}_{j \in N}$, $\mu^\varepsilon$ assigns value uniformly.
\end{itemize}
It is straightforward to verify that $\mu^\varepsilon$ is indeed a cake measure profile that is compatible with $I$; we omit the details. Then $X$ is $\mu^\varepsilon$-envy-free, so for each pair of distinct $i, j \in N$, we have
\begin{align*}
\alpha^C_i(X'_i) + \varepsilon &= \mu^\varepsilon_i(X'_i) + \varepsilon
\\ &\geq \mu^\varepsilon_i(X_i)
\\ &\geq \mu^\varepsilon_i(X_j)
\\ &\geq (1-\varepsilon) \cdot \mu^\varepsilon_i(X'_j)
\\ &= (1-\varepsilon) \cdot \alpha^C_i(X'_j)
\end{align*}
Indeed, the first inequality holds because $X_i$ cannot have more than all the value of $X'_i$ plus all of the value across reassigned pieces of the cake, the second inequality follows because $X$ is $\mu^\varepsilon$-envy-free, and the third inequality holds because $\mu^\varepsilon_i$ deems that at least $(1-\varepsilon)$ of each cell that $j$ receives under $X'_j$ is also received by $j$ under $X_j$. Since $\varepsilon > 0$ was arbitrary, thus $\alpha^C_i(X'_i) \geq \alpha^C_i(X'_j)$. Since $i, j \in N$ were arbitrary, thus $X'$ is envy-free for every cake measure profile compatible with $I$, as desired.

\vspace{\baselineskip} \noindent \textsc{Case 2:} $\mathsf{Q}(I)$ is a query. In this case, since $\mathsf{Q}$ is cellular, thus the query asks some agent $i \in N$ to cut some cake cell $S^C \in \mathcal{P}^C$ using the proportion $p \in [0, 1]$. Write $S^T \equiv \mathsf{m}(S^C)$, and let $a, b \in [0, 1]$ be such that $S^C = (a, b]$. We consider three subcases.

\vspace{\baselineskip} \textsc{Case 2a:} $p \in \{0, 1\}$. Intuitively, in this case the query is useless. In this case, the next construction state is defined by (i)~updating the cake infoset to be the mediator's next information set, which is well-defined because the current information set determines the response to a useless query whose input is a cake cell, and (ii)~leaving both appraised partitions unchanged.

\vspace{\baselineskip} \textsc{Case 2b:} $p \in (0, 1)$ and there have been fewer than $c-1$ cuts thus far, formally measured by the number of queries recorded in the chronicle of $I$. In this case, $i$ cuts the time cell $S^T$ using the \textsc{Tripartition Algorithm} and the target value $v \equiv p \cdot \alpha^C_i(S^C)$. Formally, in this case, we do the following.

First, define the {\it first time cell} $S^{T1} \equiv \mathcal{T}_i(v|S^T)$, define the {\it second time cell} $S^{T2} \equiv S^T \backslash S^{T1}$, construct $\mathcal{P}^{T+}$ from $\mathcal{P}^T$ by replacing $S^T$ with $S^{T1}$ and $S^{T2}$.

Second, define $\alpha^{T+}$ as the restriction of the given utility function profile $u$ to $\mathcal{P}^{T+}$.

Third, define the {\it first cake cell} $S^{C1} \equiv (a, \frac{a+b}{2}]$, define the {\it second cake cell} $S^{C2} \equiv (\frac{a+b}{2}, b]$, and construct $\mathcal{P}^{C+}$ from $\mathcal{P}^C$ by replacing $S^C$ with $S^{C1}$ and $S^{C2}$. For our purposes, we are free to specify that the cut point revealed by the response to the cake protocol is the midpoint of the input cake cell.

Fourth, define $\mathsf{m}^+$ by matching $S^{C1}$ to $S^{T1}$, matching $S^{C2}$ to $S^{T2}$, and preserving the matchings previously specified by $\mathsf{m}$ for all members of $\mathcal{P}^{C+} \backslash \{S^{C1}, S^{C2}\}$.

Fifth, define $\alpha^{C+}$ such that for each $i \in N$ and each $S \in \mathcal{P}^{C+}$, $\alpha^{C+}_i(S) = \alpha^{T+}_i(\mathsf{m}^+(S))$. In other words, the new cake appraisals equal the associated new time appraisals.

Finally, we define $I^+$ to be the information set that succeeds $I$ after the mediator receives the query response compatible with $(\mathcal{P}^{C+}, \alpha^{C+})$. We claim that this is well-defined, and to prove this we make an inductive argument. Indeed, by \hyperlink{Proposition1}{Proposition~1} and the partial description of the time protocol we have provided thus far, since (i)~$\mathcal{P}^T$ is constructed from $\{T\}$ by iteratively applying the Tripartition Algorithm fewer than $c-1$ times, and (ii)~for each $i \in N$ we have $u_i \in \mathcal{U}_{\mathsf{d}(c)} \cap \mathcal{U}_{\text{M}}$, thus the agents agree that each member of $\mathcal{P}^T$ is $5$-divisible, so by \hyperlink{Proposition1}{Proposition~1} each cut from the beginning through the cut that yields $\mathcal{P}^{T+}$ constructs a schedule with the given target value.\footnote{Indeed, recall the argument from Footnote 8 for Definition 10.} Moreover, since for each $i \in N$ we have $u_i \in \mathcal{U}_1 \cap \mathcal{U}_{\text{M}}$, there is consensus that each time slot has positive value and thus that each nonempty schedule has positive value; thus by induction using the definition of the time protocol, after each cut the cutter measures all time cells to have positive value, so there is consensus that all time cells have positive value. Thus there is a cake measure profile compatible with $(\mathcal{P}^{C+}, \alpha^{C+})$; for example, have each agent assign value to cake cells in accordance with the cake appraisal and distribute value within cake cells uniformly. Such a cake measure profile yields a response to the query that indeed cuts $S^C$ at its midpoint, and thus $I^+$ is well-defined; we omit the details.

\vspace{\baselineskip} \textsc{Case 2c:} $p \in (0, 1)$ and there have been exactly $c-1$ cuts thus far. This case is analogous to Case~2b, except that $i$ cuts using the \textsc{Greedy~Algorithm} instead of using the \textsc{Tripartition~Algorithm}. The only significant difference appears in the final paragraph of Case~2b: in Case~2c, the agents only agree that each member of $\mathcal{P}^T$ is $1$-divisible, so (i)~by Case~2b, each cut from the beginning through the cut that yields $\mathcal{P}^T$ constructs a schedule with the given target value, and (ii)~by \hyperlink{TheoremMK}{Theorem~MK}, the cut that yields $\mathcal{P}^{T+}$ constructs a schedule with the given target value.

\vspace{\baselineskip} \noindent \textsc{Step 3:} Conclude.

\vspace{\baselineskip} To summarize, the time protocol iteratively takes the construction state, and uses $\mathsf{Q}$ to either move to a next stage or terminate. First, if $\mathsf{Q}$ selects a useless query, then we simply move to the next stage of the cake protocol. Second, if $\mathsf{Q}$ selects a not-useless query and there have not yet been $c-1$ cuts, then the query is used to cut the time partition using the \textsc{Tripartition~Algorithm}, the new time partition is appraised using the given utility function profile, the cake partition is automatically cut by dividing the query input cell in half, and the cake appraisal is chosen to match the time appraisal; we prove in Step~2 that this is indeed compatible with moving to a next stage of the cake protocol. Third, if $\mathsf{Q}$ selects a not-useless query and there have been exactly $c-1$ cuts, then we do the same except using the \textsc{Greedy~Algorithm} instead of using the \textsc{Tripartition~Algorithm}. Finally, because $\mathsf{cost}(\mathsf{Q}) \leq c$, the only remaining case is that $\mathsf{Q}$ selects a cake allocation, in which case we modify the cake allocation to assign each agent a union of cake cells if necessary and then terminate with the associated time allocation. As argued above, the cake allocation is envy-free; thus as the time appraisal matches the cake appraisal, the time allocation is envy-free.~$\blacksquare$

\hypertarget{Appendix4}{}
\setcounter{secnumdepth}{0}
\section{Appendix 4}

In this appendix, we provide formal definitions of {\it cake protocol}, {\it cellular cake protocol}, {\it envy-free cake protocol}, and {\it cost of an envy-free cake protocol}. We do so using the recent general formalization of cake protocols in \cite{Brandl-Mackenzie2026}, adapted to simplify our arguments at no cost to our main message,\footnote{To be precise, compared to \cite{Brandl-Mackenzie2026}, we (i)~restrict attention to the interval cake, (ii)~restrict the domain of cake measures to be piecewise-uniform, and (iii)~restrict attention to protocols that select queries whose input servings are cells of the current partition. The first two strengthen our results, allowing us to apply our techniques to protocols that take advantage of the interval cake structure and that are only guaranteed to work for the smaller class of cake measures. The third change allows us to simplify our arguments by avoiding technical issues caused by non-cell queries that are addressed in \cite{Brandl-Mackenzie2026}, and comes at no cost to our main message because the protocols from the literature that we apply our techniques to satisfy this requirement after a superficial modification.} and we refer the interested reader there for further discussion and illustrative examples. It should be understood that for this appendix, all previously defined notation is free to be recycled.

\vspace{\baselineskip} \noindent \textsc{Definition 11:} Fix an economy. The associated {\it (interval) cake setting} is defined as follows.
\begin{itemize}
\item The set of {\it agents} is the same as in the given economy; we denote it $N \equiv \{1, 2, \dots, n\}$.

\item $C = (0, 1]$ is the {\it interval cake}.

\item A {\it serving} is any finite union of intervals of the form $(a, b]$, and $\mathcal{S} \subseteq 2^C$ denotes the set of servings. Observe that $\mathcal{S}$ is a Boolean algebra that includes no singletons.

\item A {\it (positive piecewise-uniform) cake measure} is a probability measure $\mu_0: \mathcal{S} \to [0, 1]$ such that there are (i)~a finite partition $\mathcal{P}$ of $C$ consisting of nonempty intervals of the form $(a, b]$, and (ii)~a list of positive convex combination weights $(w_S)_{S \in \mathcal{P}} \in (0, 1]^\mathcal{P}$ with $\sum_{S \in \mathcal{P}} w_S = 1$, such that $\mu_0$ is the associated convex combination of the associated uniform distributions.\footnote{Formally, for each $S \in \mathcal{P}$, let $\mu_0^S$ denote the uniform distribution on $S$; then $\mu_0 = \sum_{S \in \mathcal{P}} w_S \cdot \mu_0^S$.} Observe that in this case, (i)~$\mu_0$ assigns a positive value to each nonempty serving, and (ii)~for each $A \in \mathcal{S}$ and each $p \in [0, 1]$, there is $t \in [0, 1]$ such that $\mu_0(A \cap (0, t]) = p \cdot \mu_0(A)$. We let $\mathbb{M} \subseteq [0, 1]^\mathcal{S}$ denote the collection of cake measures. Observe that $\mathbb{M}$ is nonempty.

\item $\mathcal{X} \subseteq \mathcal{S}^N$ is the set of {\it cake allocations}: for each $X \in \mathcal{S}^N$, we have $X \in \mathcal{X}$ if and only if (i)~for each pair of distinct $i, j \in N$, we have $X_i \cap X_j = \emptyset$, and (ii)~$\cup_{i \in N} X_i = C$.

\item $\mathbb{D} \equiv \mathbb{M}^N$ is the {\it unrestricted domain (of all cake measure profiles)}. For each $\mu \in \mathbb{D}$, a cake allocation $X$ is {\it $\mu$-envy-free} if for each pair $i, j \in N$ we have $\mu_i(X_i) \geq \mu_i(X_j)$.
\end{itemize}
For emphasis, the utility function profile summarizing preferences in the (time division) economy is given, but the cake measure profile summarizing preferences in the associated cake setting is not known.

\vspace{\baselineskip} \noindent \textsc{Definition 12:} Fix a cake setting.
\begin{itemize}
\item A {\it query} is a tuple $\mathsf{q} = (i, S, p)$ specifying (i)~a {\it cutter} $i \in N$, (ii)~a serving $S \in \mathcal{S}$, and (iii)~a target proportion $p \in [0, 1]$. The interpretation is that $i$ is asked to cut $S$ at the leftmost point such that the proportion of the value in $S$ to the left of the cut point is $p$. We let $\mathcal{Q}$ denote the set of queries.

\item For each query $\mathsf{q} = (i, S, p)$ and each $\mu_i \in \mathbb{M}$, we define the {\it $(\mathsf{q}|\mu_i)$-threshold} by $\tau(\mathsf{q} | \mu_i) \equiv \inf \{t \in [0, 1] | \mu_i(S \cap (0, t]) \geq p \cdot \mu_i(S) \}$ and we define the {\it $(\mathsf{q}|\mu_i)$-serving} by $S_{\mathsf{q} | \mu_i} \equiv S \cap (0, \tau(\mathsf{q} | \mu_i)]$. The interpretation is that when the mediator selects query $\mathsf{q}$ and the cutter $i$ has preferences represented by $\mu_i$, the output serving is $S_{\mathsf{q} | \mu_i}$.

\item For each query $\mathsf{q} = (i, S, p)$ and each $\mu \in \mathbb{D}$, the {\it $(\mathsf{q}|\mu)$-response}, $\mathsf{r}(\mathsf{q} | \mu)$, is the pair $(\{S, S_{\mathsf{q}|\mu_i}\}, \alpha')$ such that for each $j \in N$, $\alpha'_j = \mu_j \restriction_{\{S, S_{\mathsf{q}|\mu_i}\}}$. In other words, after $i$ cuts $S$ such that $S_{\mathsf{q} | \mu_i}$ is to the left of the cut point, {\it every agent} evaluates both $S$ and $S_{\mathsf{q} | \mu_i}$ (and thereby reveals his value for $S \backslash S_{\mathsf{q} | \mu_i}$ as well). The {\it $(\mathsf{q}|\mu)$-domain}, $\mathbb{D}(\mathsf{q}|\mu)$, is the domain $\{ \mu' \in \mathbb{D} | \text{for each } j \in N,  \alpha'_j = \mu'_j \restriction_{\{S, S_{\mathsf{q} | \mu_i}\}}\}$; this is the collection of measure profiles consistent with the cut and the evaluations.
\end{itemize}
Intuitively, in a cake protocol, a mediator iteratively selects queries and receives responses to learn about the unknown measure profile.

\vspace{\baselineskip} \noindent \textsc{Definition 13:} Fix a cake setting. The {\it division game form} is the two-player game form defined as follows.
\begin{itemize}
\item The two players are the {\it mediator} and {\it nature}. There are also histories partially ordered by precedence, active players and actions associated with non-terminal histories, and information sets. The precedence order is such that each history and action at that history together determine an {\it immediate successor}, or an earliest history that follows. We define these remaining ingredients below, defining precedence entirely using immediate successors.

\item The initial history is non-terminal, its active player is nature, and its set of available actions is~$\mathbb{M}^N$. Nature's choice of measure profile is not observed by the mediator. After each choice of nature, the immediate successor is non-terminal.

\item At each non-terminal history other than the initial history, the active player is the mediator and the set of available actions is $\mathcal{Q} \cup \mathcal{X}$. We denote the set of histories where the mediator is the active player by $H$. If the mediator selects a query, then the immediate successor is non-terminal. If the mediator selects a cake allocation, then the immediate successor is terminal.

\item Each non-initial history $h$ is identified with the finite sequence of actions that leads to $h$ from the initial history: there is a {\it (query) count} $c \in \mathbb{N}_0$ such that (i)~if $h$ is non-terminal, then $h$ can be written $(\mu, \mathsf{q}_1, \mathsf{q}_2, ..., \mathsf{q}_c)$, and (ii)~if $h$ is terminal, then $h$ can be written $(\mu, \mathsf{q}_1, \mathsf{q}_2, ..., \mathsf{q}_c, X)$. In either case, (i)~the {\it chronicle at $h$} is the list of queries and their responses, $(\mathsf{q}_1, \mathsf{r}(\mathsf{q}_1|\mu), \mathsf{q}_2, \mathsf{r}(\mathsf{q}_2|\mu), ..., \mathsf{q}_c, \mathsf{r}(\mathsf{q}_c|\mu))$, and (ii)~the {\it partition at $h$} is the partition of $C$, $\mathcal{P}(h)$, generated by the input servings (of the queries) and the output servings (of the responses) in the chronicle at $h$, which by convention we take to be $\{C\}$ when $c=0$. Moreover, the {\it $h$-domain} is $\mathbb{D}(h) \equiv \cap_{t \in \{1, 2, ..., c\}} \mathbb{D}(\mathsf{q}_t|\mu)$, where by convention $c=0$ implies $\mathbb{D}(h) = \mathbb{D}$; this is nonempty as $\mu \in \mathbb{D}(h)$. Finally, two histories share an {\it information set} if and only if (i)~the mediator is the active player at both histories, and (ii)~the two histories have the same chronicle. Observe that if $h$ and $h'$ share an information set, then $\mathbb{D}(h) = \mathbb{D}(h')$ and $\mathcal{P}(h) = \mathcal{P}(h')$.

\item A play is a maximal chain of histories. If a play is finite, then it ends with a terminal history after the mediator selects a cake allocation.
\end{itemize}
Intuitively, the mediator's goal is to identify a cake allocation that is surely envy-free using as few queries as possible.

\vspace{\baselineskip} \noindent \textsc{Definition 14:} Fix a cake setting. A {\it cake (query) protocol} is a pure strategy for the mediator in the division game form. Equivalently, a cake protocol is a function $\mathsf{Q}: H \to \mathcal{Q} \cup \mathcal{X}$ such that for each pair $h, h' \in H$, if $h$ and $h'$ share an information set, then $\mathsf{Q}(h) = \mathsf{Q}(h')$. In this case, for each $\mu \in \mathbb{D}$, we define $\mathsf{cost}(\mathsf{Q}|\mu) \in \mathbb{N}_0 \cup \{\infty\}$ and $\mathsf{out}(\mathsf{Q}|\mu) \in \mathcal{X} \cup \{\mathsf{error}\}$ as follows.
\begin{itemize}
\item If the play determined by $\mu$ and $\mathsf{Q}$ is infinite, then we define $\mathsf{cost}(\mathsf{Q}|\mu) \equiv \infty$ and $\mathsf{out}(\mathsf{Q}|\mu) \equiv \mathsf{error}$.

\item If the play determined by $\mu$ and $\mathsf{Q}$ is finite, then let $c$ denote the query count of its terminal history and let $X$ denote its final action. In this case, we define $\mathsf{cost}(\mathsf{Q}|\mu) \equiv c$ and $\mathsf{out}(\mathsf{Q}|\mu) \equiv X$.
\end{itemize}
First, we say that $\mathsf{Q}$ is {\it envy-free} if for each $\mu \in \mathbb{D}$, $\mathsf{out}(\mathsf{Q}|\mu) \in \mathcal{X}$ and $\mathsf{out}(\mathsf{Q}|\mu)$ is $\mu$-envy-free. Second, we say that $\mathsf{Q}$ is {\it cellular} if for each $h \in H$ such that $\mathsf{Q}(h) \in \mathcal{Q}$, the serving of the query $\mathsf{Q}(h)$ belongs to the partition $\mathcal{P}(h)$. Finally, the {\it (worst-case) cost of $\mathsf{Q}$} is $\mathsf{cost}(\mathsf{Q}) \equiv \sup_{\mu \in \mathbb{D}} \mathsf{cost}(\mathsf{Q}|\mu)$.

\phantomsection

\end{document}